\documentclass[11pt,reprint,showpacs,showkeys,amsfonts,amssymb,amsmath,nofootinbib]{article}
\usepackage{bm,amsmath,epsf,epsfig,latexsym,amssymb,cite,lmodern}
\usepackage{graphics,psfrag}
\usepackage{hyperref}
\newcommand{\up}{{\hat{\mathbf{p}}}}
\newcommand{\vp}{{\mathbf{p}}}
\newcommand{\vq}{{\mathbf{q}}}

\newcommand{\vk}{{\mathbf{k}}}

\newcommand{\bg}{\begin{align}}
\newcommand{\eeg}{\end{align}}
\newcommand{\be}{\begin{equation}}
\newcommand{\ee}{\end{equation}}
\newcommand{\ba}{\begin{eqnarray}}
\newcommand{\ea}{\end{eqnarray}}

\newcommand{\nn}{\nonumber}

\newcommand{\ve}{\varepsilon}

\newcommand{\vz}{\mathbf{z}}

\newcommand{\vep}{\varepsilon}

\newcommand{\sth}{s_{\rm th}}

\usepackage{pdfpages,float,pstool,amssymb}
\usepackage{color,graphicx,epstopdf}
\usepackage{caption,subcaption}
\DeclareGraphicsExtensions{.ps}
\usepackage{pstool}
\usepackage{pdftricks}

\begin{document}

\thispagestyle{empty}

\title{\Large \bf A chiral covariant approach to {\boldmath$\rho\rho$} scattering}

\author{D.~G\"ulmez$^a$, U.-G.~Mei{\ss}ner$^{a,b}$, J.~A.~Oller$^c$\\[0.5em]
{\small\it  $^a$Helmholtz-Institut f\"ur Strahlen- und Kernphysik and
Bethe Center for Theoretical Physics,}\\
{\small\it Universit\"at Bonn, D--53115
Bonn, Germany}\\[0.3em]
{\small\it  $^b$Institute for Advanced Simulation, Institut f{\"u}r
Kernphysik and J\"ulich Center for Hadron Physics,}\\
{\small\it Forschungszentrum  J{\"u}lich, D-52425 J{\"u}lich, Germany}  
\\[0.3em]
{\small {\it $^c$Departamento de F\'{\i}sica, Universidad de Murcia, E-30071 Murcia, Spain}}
}

\date{}

\maketitle
\begin{abstract}
We analyze vector meson -- vector meson scattering in a unitarized chiral theory
based on a chiral covariant framework. We show that a pole assigned to the 
scalar meson $f_0(1370)$ can
be dynamically generated from the $\rho\rho$ interaction, while this is not the
case for the tensor meson $f_2(1270)$ as found in earlier works. We show that the
generation of the tensor state is untenable due to 
 the extreme non-relativistic 
kinematics used before. We further consider the effects arising from the 
coupling of channels with different orbital angular momenta which are also important. We suggest to use 
the formalism outlined here to obtain more 
reliable results for the dynamical generation of resonances in the vector-vector interaction.
\end{abstract}

\section{Introduction}
\label{sec:intro}

It is now commonly accepted that some hadron resonances are generated
by strong non-perturbative hadron-hadron interactions. Arguably the most famous
example is the $\Lambda (1405)$, that arises from the coupled-channel dynamics
of the strangeness $S=-1$ ground state octet meson-baryon channels in the vicinity of the  $\pi\Sigma$
and $K^-p$ thresholds~\cite{Dalitz:1959dn}. This resonance also has the outstanding feature 
of being actually the combination of two near poles, the so-called two-pole nature of the $\Lambda(1405)$. 
In a field-theoretic sense, one should consider this state as two particles.
This fact was predicted theoretically \cite{meissner.091016.2,jido.091016.3} and later unveiled
experimentally \cite{hylu.141016.1} (see also the discussion  in Ref.~\cite{pdg.071016.3}).  
Another example is the scalar meson $f_0(980)$ close to the $\bar KK$ threshold, 
that is often considered to arise due to the strong $S$-wave 
 interactions in the $\pi\pi$-$\bar KK$ system
with isospin zero \cite{weinstein.141016.1,jansen.141016.2,oller.091016.1}. 
A new twist was given to this field in Ref.~\cite{oset.081016.1}
where the $S$-wave vector-vector ($\rho\rho$)  interactions were investigated 
 and it was found that due to
the strong binding in certain channels, the $f_2(1270)$ and the $f_0(1370)$ mesons
could be explained as $\rho\rho$ bound states. 
 This approach offered also an explanation
why the tensor state $f_2$ is lighter than the scalar one $f_0$, as the leading order attraction
in the corresponding $\rho\rho$ channel is stronger. This work was followed up by extensions to
SU(3)~\cite{Geng:2008gx}, to account for radiative decays \cite{nagahiro.311016.1} 
  and many other works, see e.g the short review in Ref.~\cite{Oset:2012zza}. 

These results are certainly surprising and at odds with well-known features of the
strong interactions. In this 
respect it is a text-book result that the $f_2(1270)$ fits very well within a nearly ideally-mixed 
$P$-wave $q\bar{q}$ nonet comprising as well the  $a_2(1320)$, $f'_2(1525)$ and $K_s^*(1430)$ resonances
 \cite{Lichtenberg.161216.1,Koll:2000ke,Ricken:2000kf,Amsler.161216.1}. 
 Values for this mixing angle can be obtained from either the linear or quadratic mass relations 
as in Ref.~\cite{Amsler.161216.1}. Non-relativistic  quark model calculations \cite{62P.161216.1}, as well as 
with relativistic corrections \cite{64P.161216.1},
 predict that the coupling of the tensor mesons to $\gamma\gamma$ should be predominantly
 through helicity two by an $E1$ transition. 
This simple $q\bar{q}$ picture for the tensor $f_2(1270)$ resonance has been recently validated by the analyses performed 
in Ref.~\cite{Pennington.161216.1} of the high-statistics Belle data \cite{13P.161216.1,14P.161216.1}
 on $\gamma\gamma\to \pi\pi$ in both the neutral and charged pion channels. Another point of importance in support
 of the $q\bar{q}$ nature of the $f_2(1270)$ is Regge theory, since this resonance lays in a parallel linear 
exchange-degenerate Regge trajectory with a  ``universal'' slope parameter of around 1~GeV \cite{Pelaez.161216.1,Anisovich.161216.1}. 
 Masses and widths of the first resonances with increasing spin
 laying on this Regge trajectory ($\rho$, $f_2$, $\rho_3$, $f_4$) are nicely predicted \cite{Leutwyler.161216.1} 
by the dual-hadronic model of  Lovelace-Shapiro-Veneziano \cite{Lovelace.161216.1}.

 One should stress that the results of Ref.~\cite{oset.081016.1}
 were obtained based on extreme non-relativistic kinematics, $\vp_i^2/m_\rho^2\simeq 0$,
with $\vp$ the rho-meson three-momentum and $m_\rho$ the vector meson mass. 
This approximation, however,  leads to some severe simplifications:
\begin{itemize}
\item Due to the assumed threshold kinematics, the full $\rho$ propagator was 
      reduced to its scalar form, thus enabling the use of techniques already
      familiar from the pion-pion interaction \cite{oller.091016.1}. This was applied 
when considering the iteration of the interactions in the Bethe-Salpeter equation.
\item Based on the same argument, the algebra involving the spin and the
      isospin projectors of the two vector-meson states could considerably
      be simplified.
\end{itemize}
However, as $\sqrt{s_{\rm th}} = 2m_\rho = 1540$~MeV, the lighter of the bound states 
 is already quite far away from the $2\rho$ threshold. It is therefore legitimate to question the
assumptions made in Ref.~\cite{oset.081016.1}. In this work, we will reanalyze the
same reactions using a fully covariant approach. This is technically much more involved
than the formalism of the earlier works. However, as our aim is to scrutinize the
approximations made there, we stay as much as possible close to their choice of
parameters. 
 Additionally, we also consider coupled-channel scattering including 
channels with nonzero orbital angular momentum, that is, we go beyond the $S$-wave scattering 
approximation of Ref.~\cite{oset.081016.1}. 
 The inclusion of coupled channels is also important when moving away from threshold. 
 The authors of this reference only considered 
scattering in $S$-wave   because of the same type of near-threshold arguments. As will be shown, 
the near-threshold approximation is only reliable
very close to threshold.

 Our work is organized as follows: In Sec.~\ref{sec:formalism} we outline
the formalism to analyze $\rho\rho$ scattering in a covariant fashion. In particular,
we retain the full propagator structure of the $\rho$, which leads to a very different
analytic structure of the scattering amplitude compared to the extreme non-relativistic
framework. We also perform a partial-wave projection technique, that allows to perform the
unitarization of the tree-level scattering amplitudes using methods well established
in the literature. An elaborate presentation of our results is given in Sec.~\ref{sec:res},
where we also give a detailed comparison to the earlier work based on the 
non-relativistic framework. Next, we consider the effect of the coupling between channels
with different orbital angular momentum. We also improve the unitarization procedure by 
considering the first-iterated solution of the $N/D$ method in Sec.~\ref{nd.211216.1}, reinforcing 
our results obtained with the simpler unitarization method. We conclude with a summary and discussion in 
Sec.~\ref{sec:summ}. A detailed account of the
underlying projection formalism is given in App.~\ref{app.081016.1}.

\section{Formalism}
\label{sec:formalism}

The inclusion of vector mesons in a chiral effective Lagrangian can be done in a variety
of different ways, such as treating them as heavy gauge bosons, using a
tensor field formulation or generating them as hidden gauge particles of
the non-linear $\sigma$-model. All these approaches are equivalent, as shown
e.g. in the review~\cite{Meissner:1987ge}. While in principle the tensor
field formulation is preferable in the construction of chiral-invariant
building blocks, we stick here to the hidden symmetry approach as this 
was also used in  Ref.~\cite{oset.081016.1}.

To be specific, the  Lagrangian for the interactions among vector mesons is taken
 from the pure gauge-boson part of the non-linear chiral Lagrangian with 
hidden local  symmetry \cite{bando.071016.1,bando.071016.2},  
\begin{align}
\label{071016.1}
{\cal L}'=&-\frac{1}{4}\langle F_{\mu\nu}F^{\mu\nu}\rangle~.
\end{align}
Here, the symbol $\langle \ldots \rangle$ denotes the trace in  SU(2) flavor space  and the 
field strength tensor  $F_{\mu\nu}$ is
\begin{align}
\label{071016.2}
F_{\mu\nu}=&\partial_\mu V_\nu-\partial_\nu V_\mu-ig[V_\mu,V_\nu]~,
\end{align}
with the coupling constant $g=M_V/2 f_\pi$ and $f_\pi\approx 92$~MeV  \cite{pdg.071016.3}
the weak pion decay constant. The vector field $V_\mu$  is 
\begin{align}
\label{071016.3}
V_\mu=& 
\left(
\begin{array}{lll}
\frac{1}{\sqrt{2}}\rho^0 & \rho^+ \\
\rho^- & - \frac{1}{\sqrt{2}}\rho^0 
\end{array}
\right)
\end{align}
From the Lagrangian in Eq.~\eqref{071016.1} one can straightforwardly derive the interaction 
between three and four vector mesons and the corresponding vertices. The corresponding 
Lagrangians are denoted as ${\cal L}'_3$ and ${\cal L}'_4$, respectively.
The former one gives rise to $\rho\rho$ interactions through the exchange of a $\rho$ meson 
 and the latter  corresponds to purely contact 
 interactions. We did not include the $ \omega$ resonance in 
Eq.~\eqref{071016.3} since it does not contribute to the interaction part (in the isospin limit). 

Consider first the contact 
 vertices for the 4$\rho$ interaction. These  can be derived from 
Eq.~\eqref{071016.2} by keeping the terms proportional to $g^2$, leading to
\begin{align}
\label{081016.1}
{\cal L}'_4=&\frac{g^2}{2}\langle V_\mu V_\nu V^\mu V^\nu- V_\mu V^\mu V_\nu V^\nu\rangle~.
\end{align}

\begin{figure}[ht]
\psfrag{r1}{$\rho^+$}
\psfrag{r2}{$\rho^-$}
\psfrag{rp}{$\rho^+$}
\psfrag{rm}{$\rho^-$}
\psfrag{r0}{$\rho^0$}
\begin{center}
\includegraphics[width=.65\textwidth]{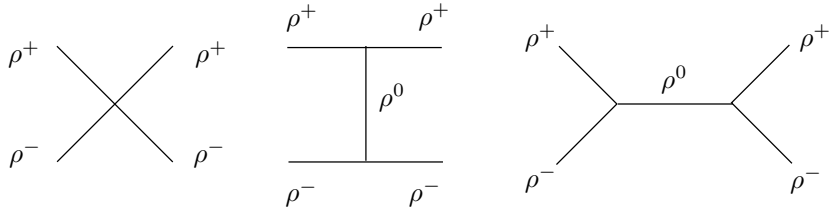}\\
\end{center}
\caption{Feynman diagrams for the tree-level amplitude $\rho^+\rho^-\to \rho^+\rho^-$.}
\label{fig.081016.1}
\end{figure}

\begin{figure}[ht]
\psfrag{r1}{$\rho^+$}
\psfrag{r2}{$\rho^-$}
\psfrag{rp}{$\rho^+$}
\psfrag{rm}{$\rho^-$}
\psfrag{r0}{$\rho^0$}
\begin{center}
\includegraphics[width=.35\textwidth]{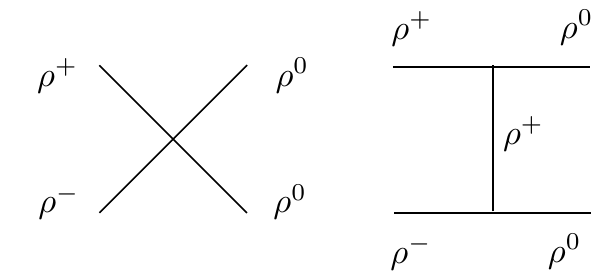}
\end{center}
\caption{Feynman diagrams for the tree-level amplitude $\rho^+\rho^-\to \rho^0\rho^0$.}
\label{fig.081016.2}
\end{figure}

The three different isospin ($I$) amplitudes for $\rho\rho$ scattering ($I=0$, 1 and $2$) 
can be worked out from the knowledge of the transitions  $\rho^+(p_1)\rho^-(p_2)\to \rho^+(p_3)\rho^-(p_4)$
and $\rho^+(p_1)\rho^-(p_2)\to \rho^0(p_3)\rho^0(p_4)$ by invoking crossing as well. We have indicated 
the different four-momenta by $p_i$, $i=1,\ldots,4$. The scattering amplitude 
for the former transition is denoted by $A(p_1,p_2,p_3,p_4)$ and the latter one by $B(p_1,p_2,p_3,p_4)$, which are shown in 
Figs.~\ref{fig.081016.1} and \ref{fig.081016.2}, respectively. 

 The contributions to those amplitudes from ${\cal L}'_4$, cf.~Eq.~\eqref{081016.1},  
 are indicated by the subscript $c$ and are given by:
\begin{align}
\label{081016.2}
A_c(k_1,k_2,k_3,k_4)=& -2 g^2(2\epsilon(1)_\mu \epsilon(2)_\nu \epsilon(3)^\nu \epsilon(4)^\mu
-\epsilon(1)_\mu\epsilon(2)^\mu\epsilon(3)_\nu\epsilon(4)^\nu
-\epsilon(1)_\mu\epsilon(2)_\nu\epsilon(3)^\mu\epsilon(4)^\nu)~,\nn\\
B_c(k_1,k_2,k_3,k_4)=& 2g^2(2\epsilon(1)_\mu \epsilon(2)^\mu \epsilon(3)_\nu \epsilon(4)^\nu 
- \epsilon(1)_\mu\epsilon(2)_\nu\epsilon(3)^\mu\epsilon(4)^\nu
- \epsilon(1)_\mu\epsilon(2)_\nu\epsilon(3)^\nu\epsilon(4)^\mu)~.
\end{align}
In this equation, the $\epsilon(i)_\mu$ corresponds to the polarization vector of the 
$i^{\rm th}$ $\rho$. Each polarization vector is characterized by its three-momentum $\vp_i$ 
and third component of the spin  $\sigma_i$ in its rest frame, 
so that $\epsilon(i)_\mu\equiv \epsilon(\vp_i,\sigma_i)_\mu$. Explicit expressions of 
these polarization vectors are given in  Eqs.~\eqref{260916.1} and \eqref{260916.3} 
of  Appendix~\ref{app.081016.1}. In the following, so as to simplify the presentation, 
the tree-level scattering amplitudes are written for real polarization vectors. 
The same expressions are valid for complex ones by taking the complex conjugate 
of the polarization vectors attached to the final particles.\footnote{The 
polarization vectors $\epsilon(\vp,\sigma)$  in the Appendix \ref{app.081016.1} are complex, so that 
the polarization vectors associated with the final-state $\rho\rho$ should be complex conjugated in this case.} 

\begin{figure}[ht]
\psfrag{rp3}{$\epsilon(\vp,\sigma_3)$}
\psfrag{rp2}{$\epsilon(\vk,\sigma_2)$}
\psfrag{rp1}{$\epsilon(\vq,\sigma_1)$}
\begin{center}
\includegraphics[width=.30005\textwidth]{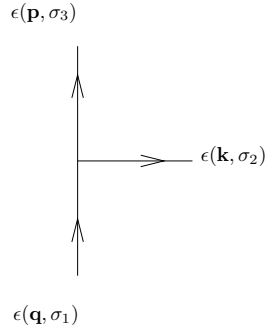}
\end{center}
\caption{Three-$\rho$ vertex from ${\cal L}'_3$.}
\label{fig.081016.3}
\end{figure} 
Considering the one-vector exchange terms, we need  the three-vector interaction Lagrangian ${\cal L}'_3$.
It  reads 
\begin{align}
\label{081016.3}
{\cal L}'_3=&ig\langle(\partial_\mu V_\nu-\partial_\nu V_\mu)V^\mu V^\nu\rangle~.
\end{align}
The basic vertex is depicted in Fig.~\ref{fig.081016.3} which after a simple calculation can be written as
\begin{align}
\label{081016.4}
V_3=&-\sqrt{2}g\Big[(q_\mu \epsilon(1)_\nu-q_\nu\epsilon(1)_\mu)\epsilon(3)^\mu \epsilon(2)^\nu
-(k_\mu \epsilon(2)_\nu-k_\nu \epsilon(2)_\mu) \epsilon(1)^{\mu} \epsilon(3)^\nu\nn\\
&-(p_\mu \epsilon(3)_\nu-p_\nu \epsilon(3)_\mu)\epsilon(2)^\mu \epsilon(1)^\nu
\Big]~.
\end{align}
In terms of this vertex, one can straightforwardly  calculate the vector exchange diagrams 
in Figs.~\ref{fig.081016.1} and 
\ref{fig.081016.2}. The expression for the $t$-channel $\rho$-exchange amplitude, the middle diagram  in Fig.~\ref{fig.081016.1}, 
and denoted by $A_t(p_1,p_2,p_3,p_4;\epsilon_1,\epsilon_2,\epsilon_3,\epsilon_4)$,  is
\begin{align}
\label{081016.6}
&A_t(p_1,p_2,p_3,p_4;\epsilon_1,\epsilon_2,\epsilon_3,\epsilon_4)=\frac{2g^2}{(p_1-p_3)^2-m_\rho^2+i 0^+} 
 \big[(p_1( p_2+ p_4) +p_3 ( p_2 + p_4)) \epsilon_1\cdot \epsilon_3 \epsilon_2\cdot \epsilon_4\nn\\
+&4(\epsilon_1\cdot k_3 \epsilon_4\cdot k_2 \epsilon_2\cdot \epsilon_3
+\epsilon_1\cdot k_3 \epsilon_2\cdot k_4 \epsilon_3\cdot\epsilon_4
+\epsilon_3\cdot k_1 \epsilon_4\cdot k_2 \epsilon_1\cdot \epsilon_2
+\epsilon_2\cdot k_4 \epsilon_3 \cdot k_1 \epsilon_1\cdot \epsilon_4
)\nn\\
-&2(\epsilon_1\cdot k_3 (\epsilon_3 \cdot k_2 + \epsilon_3 \cdot k_4) \epsilon_2\cdot\epsilon_4
+\epsilon_3\cdot k_1 (\epsilon_1\cdot k_2+\epsilon_1\cdot k_4)\epsilon_2\cdot \epsilon_4
+\epsilon_2\cdot k_4 (\epsilon_4\cdot k_1+\epsilon_4\cdot k_3)\epsilon_1\cdot \epsilon_3\nn\\
+&\epsilon_4\cdot k_2 (\epsilon_2\cdot k_1+\epsilon_2\cdot k_3)\epsilon_1\cdot\epsilon_3
)
\big]~,
\end{align}
where for short, we have rewritten $\epsilon(i)\to \epsilon_i$, and the scalar products involving 
polarization vectors  are indicated with a dot.
The $u$-channel $\rho$-exchange amplitude $A_u(p_1,p_2,p_3,p_4;\epsilon_1,\epsilon_2,\epsilon_3,\epsilon_4)$ 
can be obtained from the expression of $A_t$ by exchanging $p_3\leftrightarrow p_4$ and 
$\epsilon_3\leftrightarrow \epsilon_4$. 
In the exchange for the polarization vectors they always refer to the same arguments of 
three-momentum and spin, that is, 
$\epsilon(\vp_3,\sigma_3)\leftrightarrow\epsilon(\vp_4,\sigma_4)$. In this way,
\begin{align}
\label{081016.7}
A_u(p_1,p_2,p_3,p_4;\epsilon_1,\epsilon_2,\epsilon_3,\epsilon_4)=&
A_t(p_1,p_2,p_4,p_3;\epsilon_1,\epsilon_2,\epsilon_4,\epsilon_3)~.
\end{align}
Notice that the second diagram in  Fig.~\ref{fig.081016.2} is a sum of the $t$-channel and $u$-channel 
$\rho$-exchange diagrams.

 The $s$-channel exchange amplitude (the last diagram in Fig.~\ref{fig.081016.1}) can also be obtained from $A_t$ 
by performing the exchange $p_2\leftrightarrow -p_3$ and $\epsilon_2\leftrightarrow \epsilon_3$, with the 
same remark as above  for the exchange of polarization vectors. We then have:
\begin{align}
\label{081016.8}
A_s(p_1,p_2,p_3,p_4;\epsilon_1,\epsilon_2,\epsilon_3,\epsilon_4)=&
A_t(p_1,-p_3,-p_2,p_4;\epsilon_1,\epsilon_3,\epsilon_2,\epsilon_4)~.
\end{align}

The total amplitudes for $\rho^+\rho^-\to \rho^+\rho^-$ and $\rho^+\rho^-\to \rho^0\rho^0$ are 
\begin{align}
\label{081016.9}
A=&A_c+A_t+A_s~,\nn\\
B=&B_c+A_t+A_u~,
\end{align}
with the usual arguments $(p_1,p_2,p_3,p_4;\epsilon_1,\epsilon_2,\epsilon_3,\epsilon_4)$. 
By crossing we also obtain the amplitude for $\rho^+\rho^+\to \rho^+\rho^+$ [that we denote as 
$C(p_1,p_2,p_3,p_4;\epsilon_1,\epsilon_2,\epsilon_3,\epsilon_4)$] from the one for   $\rho^+\rho^-\to \rho^+\rho^-$ 
by exchanging $p_2\leftrightarrow -p_4$ and $\epsilon_2\leftrightarrow \epsilon_4$, that is,
\begin{align}
\label{081016.10}
C(p_1,p_2,p_3,p_4;\epsilon_1,\epsilon_2,\epsilon_3,\epsilon_4)= &
A(p_1,-p_4,p_3,-p_2;\epsilon_1,\epsilon_4,\epsilon_3,\epsilon_2)~.
\end{align}
The amplitude $C$ is purely $I=2$, that we denote as $T^{(2)}$.
 The amplitude $B$ is an admixture of the $I=0$, $T^{(0)}$, and $I=2$ amplitudes, 
\begin{align}
\label{081016.11}
B=&\frac{1}{3}(T^{(0)}-T^{(2)})~,
\end{align}
from which we find that
\begin{align}
\label{081016.12}
T^{(0)}=&3B+C~.
\end{align}
To isolate the $I=1$ amplitude, $T^{(1)}$, we take the $\rho^+\rho^-$ elastic amplitude 
$A$ which obeys the following  isospin decomposition
\begin{align}
\label{081016.13}
A=&\frac{1}{6}T^{(2)}+\frac{1}{2}T^{(1)}+\frac{1}{3}T^{(0)}~.
\end{align}
Taking into account Eqs.~\eqref{081016.12} we conclude  that
\begin{align}
\label{081016.14}
T^{(1)}=2A-2B-C~.
\end{align} 
In terms of these amplitudes with well-defined isospin the expression in Eq.~\eqref{051016.6} for 
calculating the partial-wave amplitudes in the $\ell S J I$ basis (states 
with well-defined total angular momentum $J$, total spin $S$, orbital angular momentum $\ell$ and 
isospin $I$), denoted as $T^{(JI)}_{\ell S;\bar{\ell}\bar{S}}(s)$ for the transition 
$(\bar{\ell}\bar{S}JI)\to (\ell S JI)$, simplifies to
\begin{align}
\label{081016.15}
T^{(JI)}_{\ell S;\bar{\ell}\bar{S}}(s)=\frac{Y_{\bar{\ell}}^{0}(\hat{\mathbf{z}})}{2(2J+1)}
\sum_{\scriptsize{\begin{array}{l} 
\sigma_1,\sigma_2,\bar{\sigma}_1\\
\bar{\sigma}_2,m
\end{array}
}} &\!\!\!\!\!
\int d\hat{\vp}''  \, Y_\ell^m(\vp'')^*\,
(\sigma_1\sigma_2M|s_1s_2S)(m M \bar{M}|\ell S J)(\bar{\sigma}_1\bar{\sigma}_2\bar{M}| \bar{s}_1\bar{s}_2\bar{S})
(0\bar{M}\bar{M}|\bar{\ell}\bar{S}J)\nn\\
\times & 
T^{(I)}(p_1,p_2,p_3,p_4;\epsilon_1,\epsilon_2,\epsilon_3,\epsilon_4)~,
\end{align}
with  $s$  the usual Mandelstam variable,
 $\vp_1=|\vp|\hat{\vz}$, $\vp_2=-|\vp|\hat{\vz}$, $\vp_3=\vp''$ and $\vp_4=-\vp''$, $M=\sigma_1+ \sigma_2$ and $\bar{M}=\bar{\sigma}_1+\bar{\sigma}_2$

The Mandelstam variables $t$ and $u$ for $\rho\rho$ scattering in the isospin limit 
are given by $t=-2\vp^2(1-\cos\theta)$ and $u=-2\vp^2(1+\cos\theta)$, with $\theta$ the polar angle of the final 
momentum. The denominator in $A_t$ due to the 
$\rho$ propagator, cf. Eq.~\eqref{081016.6}, vanishes for $t=m_\rho^2$ and similarly the denominator in 
$A_u$  for $u=m_\rho^2$. When performing the 
angular projection in Eq.~\eqref{081016.15} these poles give rise to a left-hand cut starting 
at the branch point $s=3m_\rho^2$. This can be easily seen by considering the integration on 
$\cos\theta$ of the fraction $1/(t-m_\rho^2+i\ve)$,
 which gives the same result both for the $t$ and the $u$ channel exchange,  
\begin{align}
\label{081016.16}
\frac{1}{2}\int_{-1}^{+1}d\cos\theta \frac{1}{-2\vp^2(1-\cos\theta)-m_\rho^2+i\ve}
=&-\frac{1}{4\vp^2}\log\left(\frac{4\vp^2+m_\rho^2}{m_\rho^2}+\frac{4\vp^2}{m_\rho^4}i\ve
\right)~,
\end{align}
with $\ve\to 0^+$. 
The argument  of the $\log$ becomes negative for $4\vp^2<-m_\rho^2$, which is equivalent to $s<3m_\rho^2$. 
Because of the  factor $\vp^2\vep$ the imaginary part of the argument of the $\log$ below the threshold 
is negative which implies that the proper value of the partial-wave amplitude on the physical axis below 
the branch point at  $s=3m_\rho^2$ is reached in the limit of vanishing negative imaginary part of $s$. 
The presence of this branch point and left-hand cut was not noticed in Ref.~\cite{oset.081016.1}, where 
only the extreme non-relativistic reduction 
was considered, so that the $\rho$ propagators in the $\rho$-exchange amplitudes collapsed to just a constant. 

Once we have calculated the partial-wave projected tree-level amplitude we proceed to its unitarization 
making use of  standard techniques within unitary chiral perturbation theory 
\cite{oller.091016.1,oller.091016.12,meissner.091016.2}. 
This is a resummation technique that restores unitarity and also allows to study the resonance region.
It has been  applied to many systems and resonances by now, e.g. in meson-meson, meson-baryon, 
nucleon-nucleon  and $WW$ systems. 
Among many others we list some pioneering works for these systems 
\cite{kaiser,oller.091016.1,guo.261016.1,meissner.091016.2,roca.091016.6,ruso.091016.8,jido.091016.3,sarkar.091016.7,oller.091016.4,oller.261016.2,delgado.091016.5,meissner.091016.9,albaladejo.091016.10,mai}. 
 In the last years  this approach has been applied also to systems containing mesons and baryons 
made from heavy quarks, some references on this topic are 
\cite{gamermann.101016.1,aceti.101610.2,romanets.101016.3,kang.101016.4,Roca:2015tea}.

The basic equation to obtain the final unitarized $T$ matrix in the subspace of coupled channels 
$\ell S J I$,  with  the same $JI$,  is \footnote{In order 
to easier the comparison with Ref.~\cite{oset.081016.1} we take the same sign convention for 
matrices $V(s)$ and $T(s)$ as in that reference.}
\begin{align}
\label{081016.17}
T^{(JI)}(s)=&\left[I-V^{(JI)}(s)\cdot G(s)\right]^{-1}\cdot V^{(JI)}(s)~.
\end{align} 
Here, $G(s)$ is a  diagonal matrix  made up by the two-point loop function $g(s)$ with $\rho\rho$ 
as intermediate states,
\begin{align}
\label{081016.18}
g(s)\to &i\int \frac{d^4 q}{(2\pi)^4}\frac{1}{(q^2-m_\rho^2)((P-q)^2-m_\rho^2)}~,
\end{align}
 where $P^2=s$ and 
within our normalization, cf. Eq.~\eqref{051016.11}, ${\rm Im}~g(s)=-|\vp|/8\pi\sqrt{s}$. 
The loop function  $g(s)$ is logarithmically divergent and it can be calculated once its value 
at a given  reference point is subtracted. In this way, one can write down a once-subtracted dispersion 
relation for $g(s)$ whose result is\footnote{It is the same result as calculating $g(s)$ in 
dimensional regularization, $d=4+2\epsilon$, and replacing the 
$1/\epsilon$ divergence by a constant, cf.~\cite{oller.091016.11}.}
\begin{align}
\label{081016.19}
g(s)=&\frac{1}{(4\pi)^2}\left(
a(\mu )+\log\frac{m^2_\rho}{\mu^2}+\sigma\left[\log(\sigma+1)-\log(\sigma-1)\right]
\right)~,
\end{align}
with
\begin{align}
\label{081016.20}
\sigma=&\sqrt{1-\frac{4m_\rho^2}{s}}~,
\end{align}
and $\mu$ is a renormalization scale typically taken around $m_\rho$, such the 
sum $a(\mu )+\log m_\rho^2/\mu^2$ is independent of $\mu$. 
  The subtraction constant in Eq.~\eqref{081016.19} could depend on the quantum numbers 
$\ell$, $S$ and $J$, but not on $I$ due to the isospin symmetry \cite{jido.091016.3}.

To compare with the results of Ref.~\cite{oset.081016.1}, we also evaluate the function $g(s)$ 
introducing a three-momentum cutoff $q_{\rm max}$,  the resulting $g(s)$ function is denoted by 
$g_c(s)$,
\begin{align}
\label{081016.21}
g_c(s) &=\frac{1}{2\pi^2}\int_0^{q_{\rm max}}dq \frac{q^2}{w(s-4w^2+i\varepsilon)}~,
\end{align}
with $w=\sqrt{q^2+m_\rho^2}$.  This integral can be done algebraically \cite{oller.091016.11}
\begin{align}
\label{081016.22}
g_c(s)=&\frac{1}{(4\pi)^2}
\left(\sigma \left[ \log\left(\sigma \sqrt{1+\frac{m_\rho^2}{q_{\rm max}^2} }+1\right)
-  \log\left(\sigma \sqrt{1+\frac{m_\rho^2}{q_{\rm max}^2} }-1\right) \right] \right.\nonumber\\
&\left. \qquad  +2\log\left\{ \frac{m_\rho}{q_{\rm max}}
\left(1+\sqrt{1+\frac{m_\rho^2}{q_{\rm max}^2}}\right)\right\}\right)~.
\end{align}
Typical values of the cutoff are around 1~GeV. 
The unitarity  loop function $g(s)$ has a branch point at the $\rho\rho$ threshold 
($s=4m_\rho^2$) and a unitarity cut above it ($s>4m_\rho^2$). The physical values of the $T$-matrix 
$T^{(JI)}(s)$, with $s>4m_\rho^2$, are
 reached in the limit of vanishing positive imaginary part of $s$. 
Notice that the left-hand cut present in $V^{(JI)}(s)$ for $s<3m_\rho^2$ 
does not overlap with the unitarity cut, 
so that  $V^{(JI)}(s)$ is analytic in the complex $s$-plane
 around the physical $s$-axis for physical energies.
In this way, the sign of the vanishing imaginary part of $s$ for $V^{(JI)}(s)$ is of 
no relevance in the prescription stated above for reaching its value  on the real 
axis with $s<3m_\rho^2$ according to the Feynman rules. 

We can also get a natural value for the subtraction constant $a$ in Eq.~\eqref{081016.19} by matching $g(s)$ and 
$g_c(s)$ at threshold where $\sigma=0$. For $\mu=m_\rho$, a usual choice, the final expression simplifies to 
\begin{align}
\label{100816.1}
a=-2\log\frac{q_{\rm max}}{m_\rho}\left(1+\sqrt{1+\frac{m_\rho^2}{q_{\rm max}^2}}\right)~.
\end{align}

It is also worth noticing that Eq.~\eqref{081016.17} gives rise to a $T$-matrix 
$T^{(IJ)}(s)$ that is gauge invariance 
in the hidden local symmetry theory because this equation just stems from the partial-wave projection 
of a complete on-shell tree-level calculation within that theory, which certainly is gauge invariant.

\section{Results}
\label{sec:res}

One of our aims  is to check the stability of the results of Ref.~\cite{oset.081016.1} 
under relativistic corrections, particularly regarding the generation of the poles that could be associated with 
the $f_0(1370)$ and $f_2(1270)$ resonances as obtained in that paper. The main source of difference between 
our calculated $V^{(JI)}(s)$ and those in Ref.~\cite{oset.081016.1} arises from the different treatment of the $\rho$-meson propagator. 
The point is that the authors of Ref.~\cite{oset.081016.1} take the non-relativistic limit of this propagator so that 
from the expression $1/(t-m_\rho^2)$, cf.~Eq.~\eqref{081016.6}, or $1/(u-m_\rho^2)$, only $-1/m_\rho^2$ is kept. 
 This is the reason  that the tree-level amplitudes calculated in Ref.~\cite{oset.081016.1} 
do not have the branch point singularity at $s=3m_\rho^2$ nor the corresponding left-hand cut for $s<3m_\rho^2$. It turns out that 
for the isoscalar tensor case, the resonance $f_2(1270)$ is below this branch point, so that its influence cannot be neglected 
when considering the generation of this pole within this approach.

\subsection{Uncoupled $S$-wave scattering}
\label{uss}

\begin{center}
 \begin{figure}[H]
 
 \begin{subfigure}{.48\textwidth}
 \includegraphics[width=\linewidth]{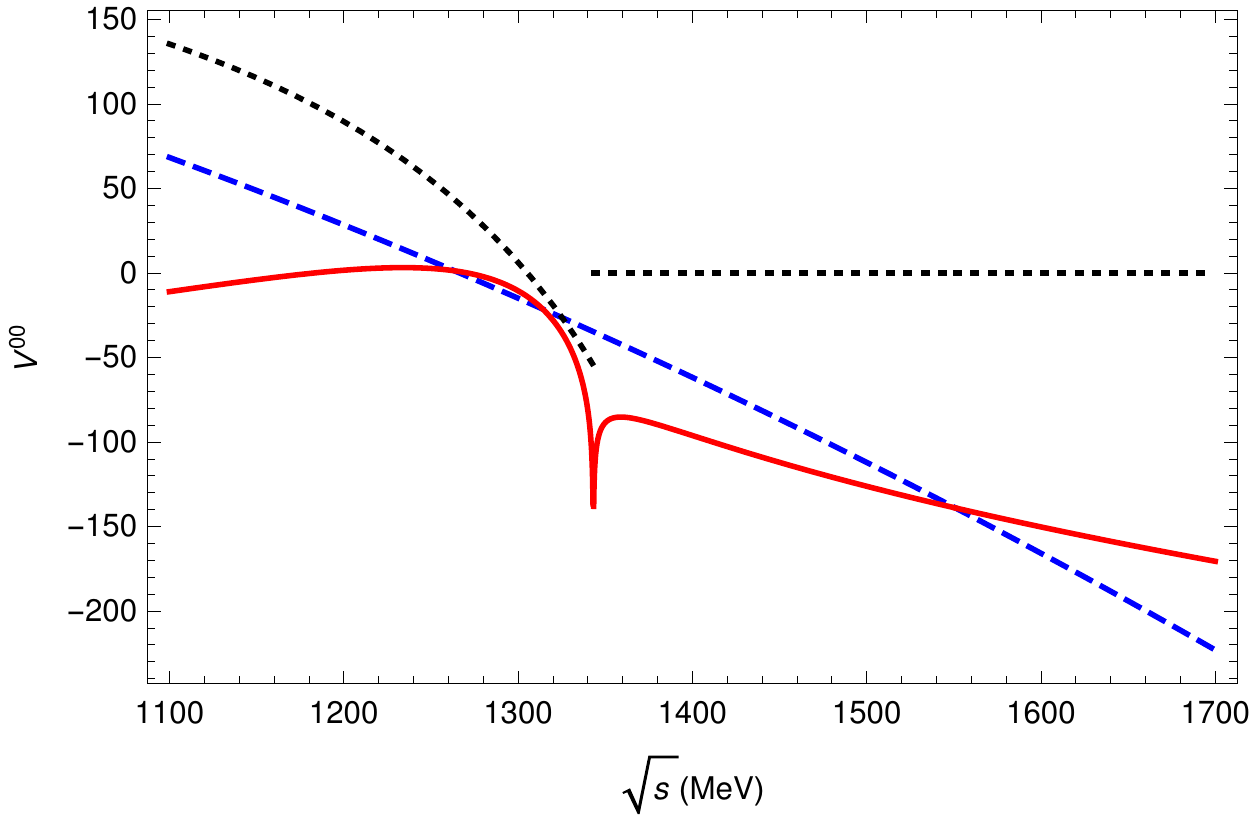}
 \caption{$V^{(00)}(s)$}
 \label{V00}
 \end{subfigure}
 \begin{subfigure}{.48\textwidth}
 \includegraphics[width=\linewidth]{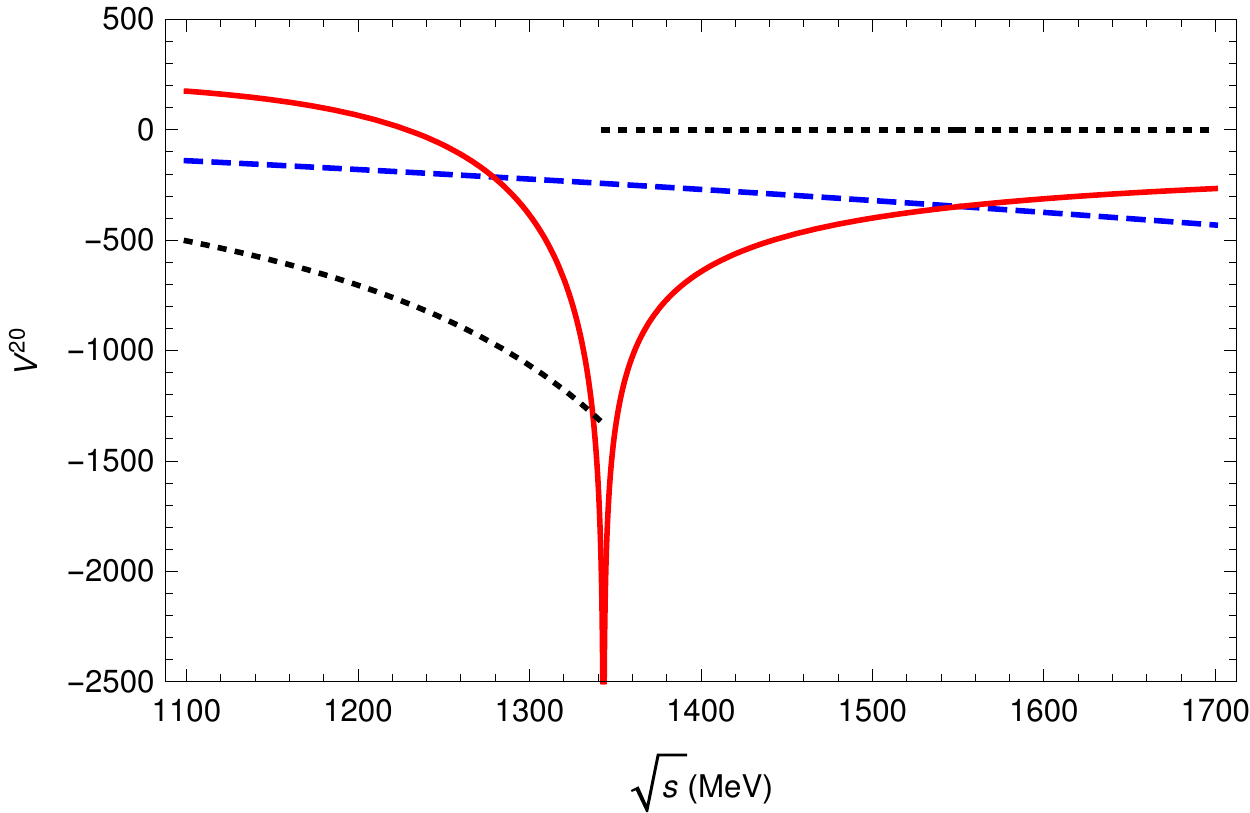}
 \caption{$V^{(20)}(s)$}
 \label{V20}
 \end{subfigure}
 \medskip
 \begin{subfigure}{.48\textwidth}
 \includegraphics[width=\linewidth]{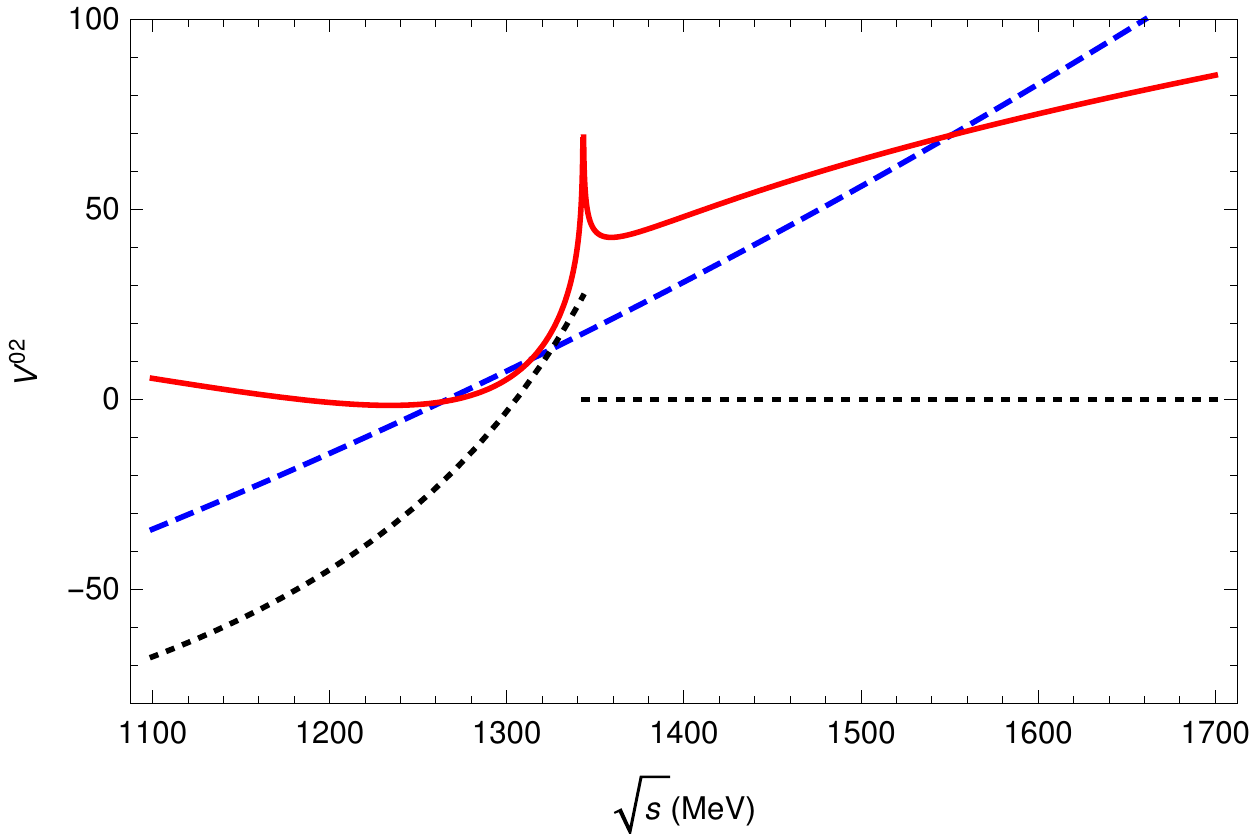}
 \caption{$V^{(02)}(s)$}
 \label{V02}
 \end{subfigure}
 \begin{subfigure}{.48\textwidth}
 \includegraphics[width=\linewidth]{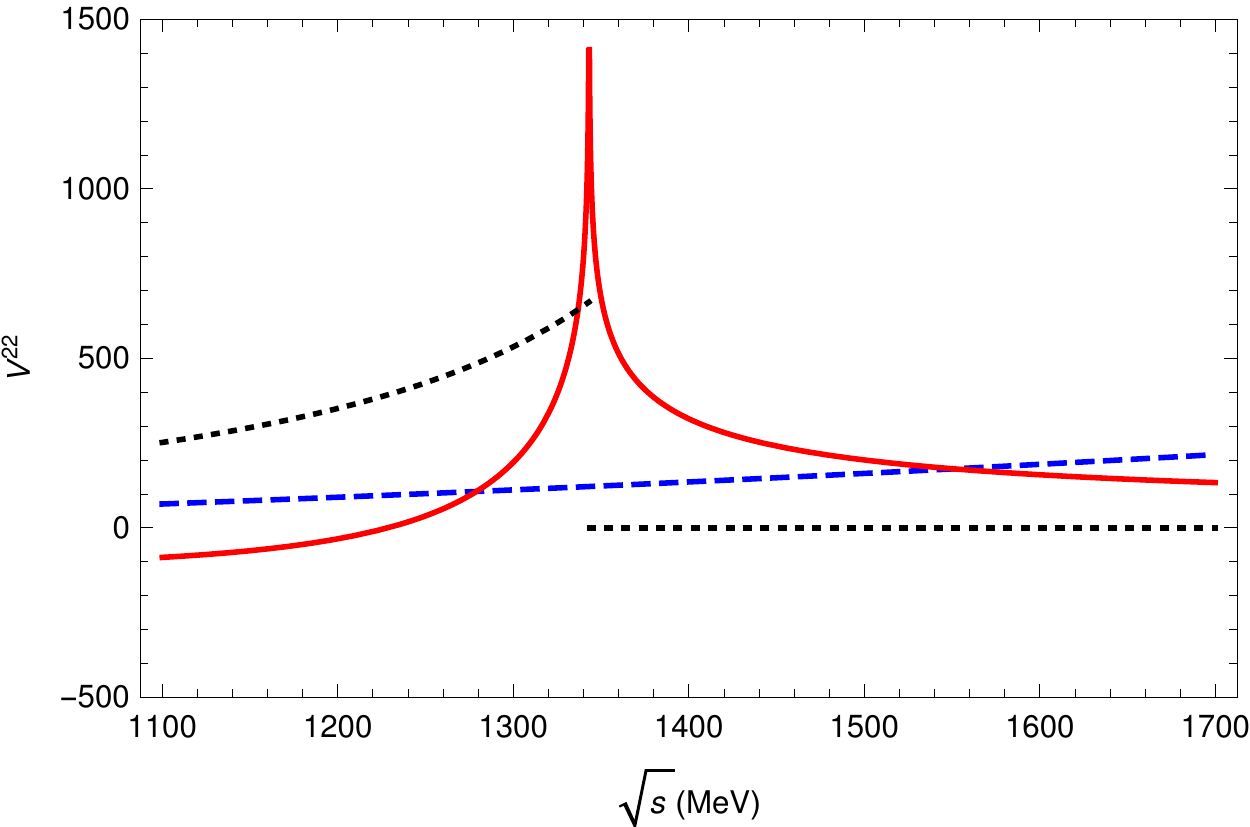}
 \caption{$V^{(22)}(s)$}
 \label{V22}
 \end{subfigure}
 \medskip
 \centering
 \begin{subfigure}{.48\textwidth}
 \includegraphics[width=\linewidth]{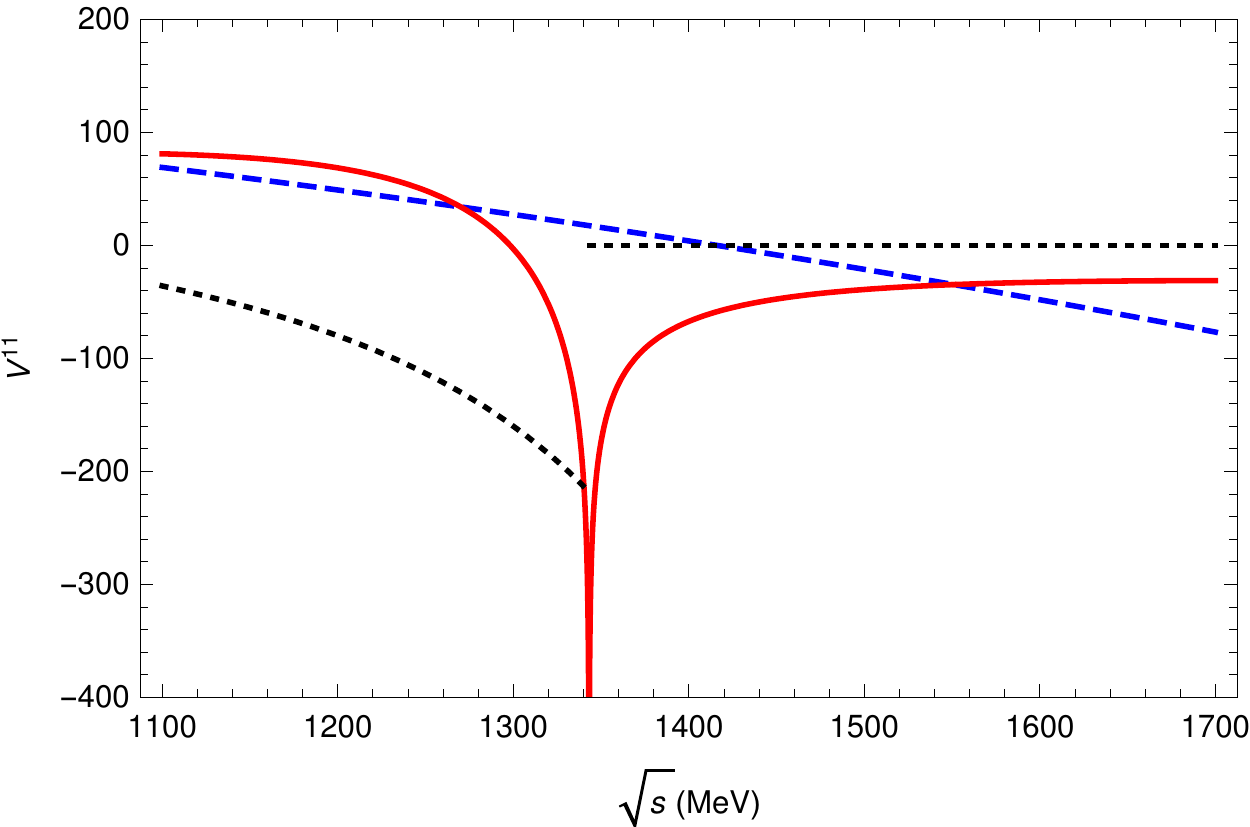}
 \caption{$V^{(11)}(s)$}
 \label{V11}
 \end{subfigure} 
 \caption{$S$-wave potentials $V^{(JI)}(s)$ (in MeV) for our 
  calculation (real part: red solid line, imaginary part: black dotted  line) 
  and for the calculation of Ref.~\cite{oset.081016.1} (blue dashed lines).}
 \label{fig:potentials}
 \end{figure}
\end{center}

 The issue on the relevance of this branch point singularity in the $\rho$-exchange amplitudes  was not  addressed in Ref.~\cite{oset.081016.1} 
and it is indeed very important.
This is illustrated in Fig.~\ref{fig:potentials} where we plot the potentials $V^{(JI)}(s)$ in $S$-wave ($\ell=0$) (only $S$-wave  scattering is 
considered in Ref.~\cite{oset.081016.1}).\footnote{Partial waves with $\ell\neq 0$ are considered in Sec.~\ref{css}.}
From top to bottom and left to right we show in the figure the potentials for the quantum numbers $(J,I)$ 
equal to $(0,0)$,  $(2,0)$,  $(0,2)$, 
 $(2,2)$ and $(1,1)$. 
  The red solid  and  black dotted lines correspond to the real and imaginary parts of our full covariant calculation  of the $V^{(JI)}(s)$, respectively, 
 while the  blue dashed ones are the results of Ref.~\cite{oset.081016.1}. 
 The imaginary part in our results for $V^{(JI)}(s)$ 
appears below $s<3m_\rho^2$ due to the left-hand cut that arises from the $t$- and $u$-channel $\rho$-exchanges.

It can be seen that our results and those of Ref.~\cite{oset.081016.1} are typically close near threshold ($s=4m_\rho^2$) but 
for lower values of $s$ they typically depart quickly due to the onset of the branch point singularity at $s=3m_\rho^2$. The strength of this 
singularity depends on the channel, being particularly noticeable in the $(J,I)=(2,0)$ channel, while for the $(0,0)$ channel 
it is comparatively weaker.

The strongest attractive potentials in the near threshold region  occur for $(J,I)=(0,0)$ and $(2,0)$ and in every of these channels Ref.~\cite{oset.081016.1} 
found a bound-state pole that the authors associated with the $f_0(1370)$ and $f_2(1272)$ resonances, respectively.
 For the $(0,0)$ quantum numbers the pole position is relatively close to the $\rho\rho$ threshold, while for $(2,0)$  
it is much further away. Two typical values of the cutoff $q_{\rm max}$ were used in Ref.~\cite{oset.081016.1}, $q_{\rm max}=875$~MeV and 
$1000$~MeV. We  employ these values here, too, together with $q_{\rm max}=m_\rho$ (so that we consider three values of $q_{\rm max}$ 
separated by around 100~MeV), and study the pole positions for our $T^{(JI)}(s)$ amplitudes in $S$ wave. We only find a bound 
state for the isoscalar scalar case, while for the tensor case no bound state is found. In Table~\ref{Polegc} we give 
the values of the pole positions for our full calculation for $q_{\rm max}=m_\rho$ (first), $875$ (second) and $1000$~MeV (third 
row). For comparison we also give in round brackets the bound state masses  obtained in Ref.~\cite{oset.081016.1}, when appropriate. 
As indicated above, the strong differences  for $V^{(20)}(s)$ between our full covariant calculation and the one in 
Ref.~\cite{oset.081016.1} in the extreme non-relativistic limit, cf. Fig.~\ref{V20}, imply the final disappearance 
of the deep bound state for the isoscalar tensor case. The nominal three-momentum of a $\rho$ around the mass of the 
$f_2(1270)$ has a modulus  of about $0.6m_\rho \simeq 460$~MeV  and for such high values of three-momentum relativistic corrections 
are of importance, as explicitly calculated here. On the contrary, the $(0,0)$ pole is located closer to the $\rho\rho$ threshold 
and the results are more stable against relativistic corrections, though one still finds differences of around 20~MeV in the bound state mass.

\begin{table}[t!]
\begin{center}
\begin{tabular}{|c|c|c|c|} \hline
$q_{\text{max}}$ (MeV) & Pole Position~(MeV)  (Mass in Ref.~\cite{oset.081016.1}) &  $(\gamma^{(00)}_{00})^2$~(GeV$^2$)  & $X^{(00)}_{00}$ \\ \hline 
$775$ & $1515.9$   & $55$ & $0.73$ \\ \hline
$875$ & $1494.8$ (1512)  & $64$ & $0.63$ \\ \hline
$1000$ & $1467.2$ (1491) & $68$ & $0.52$ \\ \hline
$775$ (conv.)& $1521.9$ &  &   \\ \hline
$875$ (conv.)& $1501.6$ &  &   \\ \hline
$1000$ (conv.)& $1475.6$ &  &   \\ \hline
\end{tabular}
\caption{Pole position for the partial wave $T^{(00)}(s)$ (2nd column), residue (3rd column) and compositeness (4th column)
 as a function of the three-momentum cutoff $q_{max}$ (1st column). 
 In the last three rows  we take into account the finite width of the $\rho$ in the evaluation of the $g(s)$ function, as indicated between parenthesis by (conv). 
 For details, see the text.}
\label{Polegc}
\end{center}
\end{table}

  In addition we also show in the third column of Table~\ref{Polegc}  the residue  of $T^{(00)}(s)$ at the pole 
position $s_P$. For a generic partial wave  $T^{(JI)}_{\ell S;\bar{\ell}S}(s)$, 
 its residue at a pole  is denoted by  $\gamma^{(JI)}_{\ell S}\gamma^{(JI)}_{\bar{\ell}\bar{S}}$ and is defined as
\begin{align}
\label{101016.2}
\gamma^{(JI)}_{\ell S} \gamma^{(JI)}_{\bar{\ell}\bar{S}} = & 
-\lim_{s\to s_P} (s-s_P) 
T^{(JI)}_{\ell S;\bar{\ell}\bar{S}}(s)~.
\end{align}
In terms of these couplings one can also calculate the compositeness $X^{(JI)}_{\ell S}$ associated with this bound state  
\cite{hyodo.101016.5,aceti.101016.6,sekihara.101016.7}, 
\begin{align}
\label{101016.3}
X_{\ell S}^{(JI)}=&-{\gamma^{(JI)}_{\ell S;\ell S}}^2 \left.\frac{\partial g(s)}{\partial s}\right|_{s_P}~,
\end{align}
which in our case determines the $\rho\rho$ component in such bound state.
 Notice that the derivative of $g(s)$ from Eq.~\eqref{081016.19} (which is negative below threshold) does not depend on the subtraction constant, 
 the dependence on the latter enters implicitly  by the actual value of the pole position $s_P$. 
Of course, if one uses a three-momentum cutoff then $g_c(s)$ must be employed in the evaluation of $X^{(JI)}_{\ell S}$. 
 The compositeness obtained for the pole positions in Table~\ref{Polegc} is  given in the fourth column  of the same table.
 As expected the $\rho\rho$ component is dominant, with $X_{00}^{(00)}>0.5$, and increases as the 
pole moves closer to threshold, so that it is 73\% for $q_{\rm max}=m_\rho$ and $\sqrt{s_P}=1516$~MeV. 

\begin{table}[t]
\begin{center}
\begin{tabular}{|c|c|c|c|} \hline
$a$ & Pole Position (MeV) &  $(\gamma^{(00)}_{00})^2$ (GeV$^2$)  & $X^{(00)}_{00}$ \\ \hline \hline
$-1.697$ & $1525.7$ & $43$ & $0.80$ \\ \hline
$-1.938$ & $1500.4$ & $56$ & $0.69$ \\ \hline
$-2.144$ & $1474.5$ & $62$ & $0.58$ \\ \hline
$-1.697$ (conv.)& $1546.0$ &  &   \\ \hline
$-1.938$ (conv.)& $1517.7$ &  &   \\ \hline
$-2.144$ (conv.)& $1491.1$ &  &   \\ \hline
\end{tabular}
\caption{Pole positions for the partial wave $T^{(00)}(s)$ (2nd column), residue (3rd column) and compositeness (4th column) as a function of 
the subtraction constant $a$ (1st column). 
 In the last three rows we take into account the finite width of the $\rho$ in the evaluation of the $g(s)$ function,
 this is indicated between parenthesis by (conv). }
\label{Poleg}
\end{center}
\end{table}

We can also determine the pole positions when $g(s)$ is calculated with exact analytical properties, Eq.~\eqref{081016.19}, and taking for $a$ 
the values from Eq.~\eqref{100816.1} as a function of $q_{\rm max}$. 
 The results are given Table~\ref{Poleg}, where we also give the residue at the pole 
position and the calculated compositeness, in the same order as in Table.~\ref{Polegc}. The results obtained are quite close to those in this table 
so that we refrain of further commenting on them.
 Nonetheless, we should stress again that we do not find any pole for the isoscalar tensor case. 

\begin{figure}[t]
\begin{center}
\includegraphics[width=0.65\textwidth]{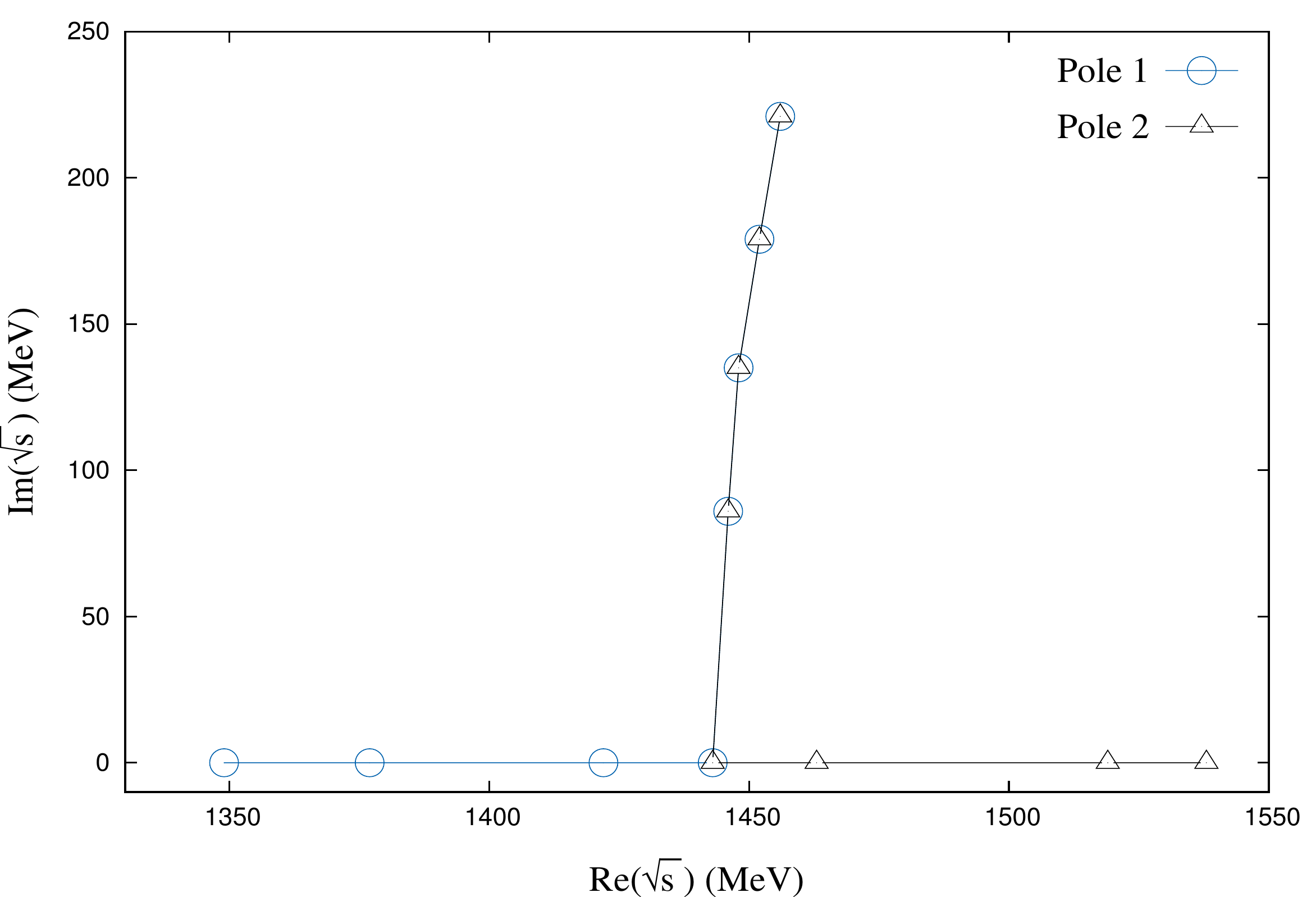}
\end{center}
\caption{Evolution of the poles in the physical Riemann sheet for  the isoscalar tensor channel as a function of $q_{\rm max}$. Two poles are present 
on the real axis for our starting value of $q_{\text{max}}=300$~MeV, 
 and along the trajectory we increase $q_{\rm max}$ in steps of $\delta q_{\text{max}}=50$ MeV. The two poles  merge at $q_{\text{max}}=403.1$~MeV 
 and for larger values of the cutoff there is one pole that moves  deeper into the complex plane.}
\label{fig:s2pole}
\end{figure}

 We could try to enforce the generation of  an isoscalar tensor pole by varying $q_{\rm max}$, when using $g_c(s)$, or by varying 
 $a$, if Eq.~\eqref{081016.19} is used. 
In the former case a much lower value of $q_{\rm max}$ is required than the chiral expansion scale around 1~GeV ($q_{\text{max}}\lesssim 400$~MeV),
 while for the latter a qualitatively similar situation arises when taking into account the relationship between $a$ and $q_{\rm max}$ of Eq.~\eqref{100816.1}.
  Even more serious are two facts that happen in relation with this isoscalar tensor pole. 
First one should stress that  such pole appears associated to the evolution with $q_{\rm max}$ or $a$ 
of a pole in the first Riemann sheet, which violates analyticity. This is shown in Fig.~\ref{fig:s2pole} where we exhibit the evolution  of this pole as a function 
of $q_{\rm max}$. We start the series at a low value of $q_{\rm max}=300$~MeV, where we have two poles on the real axis, and increase the cutoff 
in steps of $\delta q_{\rm max}=50$~MeV. 
 These two poles  get closer and merge for $q_{\rm max}=403.1$~MeV. For larger values of the cutoff the resulting pole moves deeper into the complex plane of the physical or first Riemann sheet.
 Second, we obtain that $X^{(20)}_{02}$ is larger than 1. For example,  for $q_{\text{max}}=400$~MeV, there are two poles at 1422.4 and 1463.4~MeV 
with $X^{(20)}_{02}=2.7$ and $3.8$, in order,  which of course makes no sense as compositeness factors have to be less or equal to one.

Next, we take into account the finite width of the $\rho$ meson in the evaluation of the unitarity two-point loop function $g(s)$. 
  As a result the peak in the modulus squared of the isoscalar scalar amplitude  now  acquires some width due to the width itself of 
the $\rho$ meson. To take that into account this effect we convolute the $g(s)$ function with a Lorentzian mass squared distribution for each of the two 
$\rho$ mesons in the intermediate state \cite{oset.081016.1,ruso.091016.8}. The resulting unitarity loop function is denoted by $\mathfrak{g}(s)$ and 
is given by
\begin{align}
\label{101016.4}
\mathfrak{g}(s)=&\frac{1}{N^2}\int_{(m_\rho-2\Gamma_\rho)^2}^{(m_\rho+2\Gamma_\rho)^2}dm_1^2\frac{\Gamma m_1/\pi}{(m_1^2-m_\rho^2)^2+m_1^2\Gamma^2}
\int_{(m_\rho-2\Gamma_\rho)^2}^{(m_\rho+2\Gamma_\rho)^2}dm_2^2\frac{\Gamma m_2/\pi}{(m_2^2-m_\rho^2)^2+m_2^2\Gamma^2}
g(s,m_1^2,m_2^2)~.
\end{align}
The normalization factor $N$ is 
\begin{align}
\label{101016.5}
N=\int_{(m_\rho-2\Gamma_\rho)^2}^{(m_\rho+2\Gamma_\rho)^2}dm^2\frac{\Gamma m/\pi}{(m^2-m_\rho^2)^2+m^2\Gamma^2}~,
\end{align}
with $\Gamma(m)$ the width of the $\rho$ meson with mass $m$. Due to the $P$-wave nature of this decay to $\pi\pi$, we take into account 
its strong cubic dependence on the decaying pion three-momentum and use the approximation
\begin{align}
\label{101016.6}
\Gamma(m)=&\Gamma_\rho \left(\frac{m^2-4m_\pi^2}{m_\rho^2-4m_\pi^2}\right)^3 \theta(m-2m_\pi)
\end{align} 
with $m_\pi$ the pion mass and $\Gamma_\rho\cong 148$~MeV \cite{pdg.071016.3}. 
The function $g(s,m_1^2,m_2^2)$ is the  two-point loop function with 
different masses, while in Eq.~\eqref{081016.19} we give its expression for the equal mass case.  When evaluated in terms of a dispersion 
relation it reads,
\begin{align}
\label{101016.7}
g(s,m_1^2,m_2^2)=&\frac{1}{16\pi^2}\bigg\{{a}(\mu)+\log\frac{m_1^2}{\mu^2}
+\frac{s-m_1^2+m_2^2}{2s}\log\frac{m_2^2}{m_1^2}\nonumber\\
&+\frac{\lambda^{1/2}(s)}{2s}\bigg[\log\big(\lambda^{1/2}(s)+s-m_2^2+m_1^2\big)-\log\big(\lambda^{1/2}(s)-s+m_2^2-m_1^2\big)\nonumber\\
&+\log\big(\lambda^{1/2}(s)+s+m_2^2-m_1^2\big)-\log\big(\lambda^{1/2}(s)-s-m_2^2+m_1^2\big)\bigg]\bigg\}\ ,
\end{align}
with $\lambda^{1/2}(s)=\sqrt{s^2 + m_1^4 + m_2^4 -2 s m_1^2 -2 s m_2^2 -2 m_1^2 m_2^2}$.
 The algebraic expression of this function when calculated with a three-momentum cutoff for different masses can be found in Ref.~\cite{oller.091016.11}, 
 to which  we refer the interested reader.

 \begin{figure}[t]
\begin{center}
\psfrag{AA}{$|T^{00}|^2$}
 \includegraphics[width=0.6\linewidth]{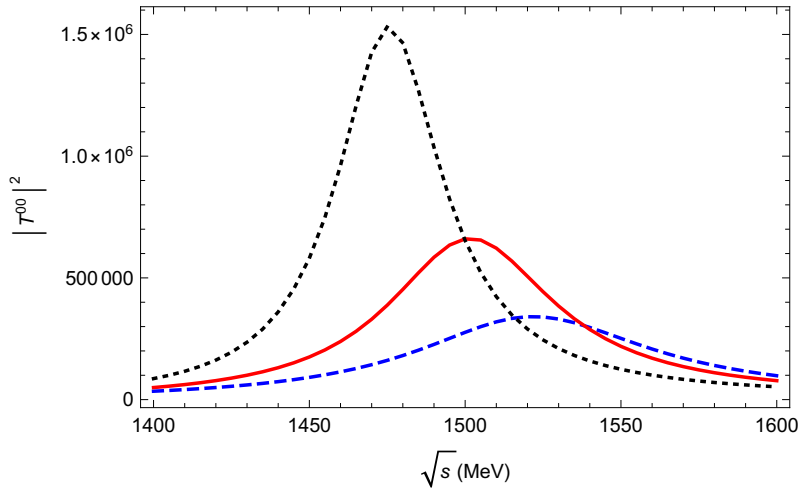}
 \caption{The amplitude squared $|T^{00}|^2$ when using the convoluted $g(s)$ function with a cutoff $q_{\rm max}$. 
 The blue dashed line corresponds to $q_{\rm max}=775$~MeV, the red solid one to $875$~MeV and the black dotted one to 1000~MeV.}
 \label{fig.211016.1}
\end{center}
 \end{figure}

When using the convoluted $g(s)$ function we find similar masses for the peak of $|T^{(00)}|^2$ in the $(0,0)$ channel compared to the case 
without convolution. 
The resulting peak positions are given in the last three rows of Tables~\ref{Polegc} and \ref{Poleg}. 
 The effects of the non-zero $\rho$ width are clearly seen in Fig.~\ref{fig.211016.1}, where we 
 plot $|T^{(00)}(s)|^2$ for the 
different values of $q_{\max}$ shown in Table~\ref{Polegc}. 
The shape of the peaks follows quite closely a Breit-Wigner form, though it is slightly wider to the right side of the peak. 
 We find that the width decreases with the increasing value of $q_{\max}$, being around 45, 65 and 95~MeV for $q_{\max}=1000$, $875$ and 
775~MeV, respectively, of similar size as those found in Ref.~\cite{oset.081016.1}.
 When using a subtraction constant instead of $q_{\rm max}$, relating them through Eq.~\eqref{100816.1},  
the picture is quite similar. The peak positions are given in the last three columns of Table~\ref{Poleg} while  the widths obtained are 
around 105, 70 and 50~MeV for $a=-1.70$, $-1.94$ and $-2.14$, in order. 
   These widths are significantly smaller than the PDG values assigned to the $f_0(1370)$ resonance  of 200-500~MeV \cite{pdg.071016.3}.

Due to the coupling of the $\rho\rho$ and $\pi \pi$, this pole could develop a 
larger width.  This is approximated in Ref.~\cite{oset.081016.1} by considering the 
imaginary part of the $\pi\pi$ box diagram, with a $\rho\to \pi\pi$ 
vertex at each of the vertices of the box. These vertices are also worked out from 
the non-linear chiral Lagrangian with 
hidden gauge symmetry \cite{bando.071016.1,bando.071016.2}.   
We refer to Ref.~\cite{oset.081016.1} for details on the
 calculation of this contribution. According to this reference one has to add 
to $V^{(00)}_{00;00}$ and to $V^{(20)}_{02;02}$ the contribution $V_{2\pi}^{(JI)}$,  given by
\begin{align}
\label{111016.1}
V_{2\pi}^{(00)}=&20i\,{\rm Im}\widetilde{V}_{\pi\pi}~,\nn\\
V_{2\pi}^{(20)}=&8i\,{\rm Im}\widetilde{V}_{\pi\pi}~.
\end{align} 
In the calculation of the function $\widetilde{V}_{\pi\pi}$, Ref.~\cite{oset.081016.1} 
introduces a monopole form factor $F(q)$ for each of the 
four $\rho\to\pi\pi$ vertices in the pion box calculation,
\begin{align}
\label{111016.2}
F(q)=\frac{\Lambda^2-m_\pi^2}{\Lambda^2-(k-q)^2}
\end{align}
with $k^0=\sqrt{s}/2$, $\vk=0$, $q^0=\sqrt{s}/2$ and $\vq$ the integration variables. 
This introduces a sizeable dependence of the results on the value of $\Lambda$. 
Nonetheless, in order to compare with Ref.~\cite{oset.081016.1} we follow the very same 
scheme of calculation and take the same values for $\Lambda$, that is, 
1200, 1300 and 1400~MeV.\footnote{Another more complete scheme is two 
work explicitly with coupled-channel scattering as done in Ref.~\cite{albaladejo.101016.1}, 
where $\rho\rho$ and $\pi\pi$ channels, among many others, were explicitly included. 
In this way resonances develop decay widths in a full nonperturbative fashion because of 
the coupling between channels.}

\begin{center}
 \begin{figure}
\begin{tabular}{cc}
 \includegraphics[width=0.48\textwidth]{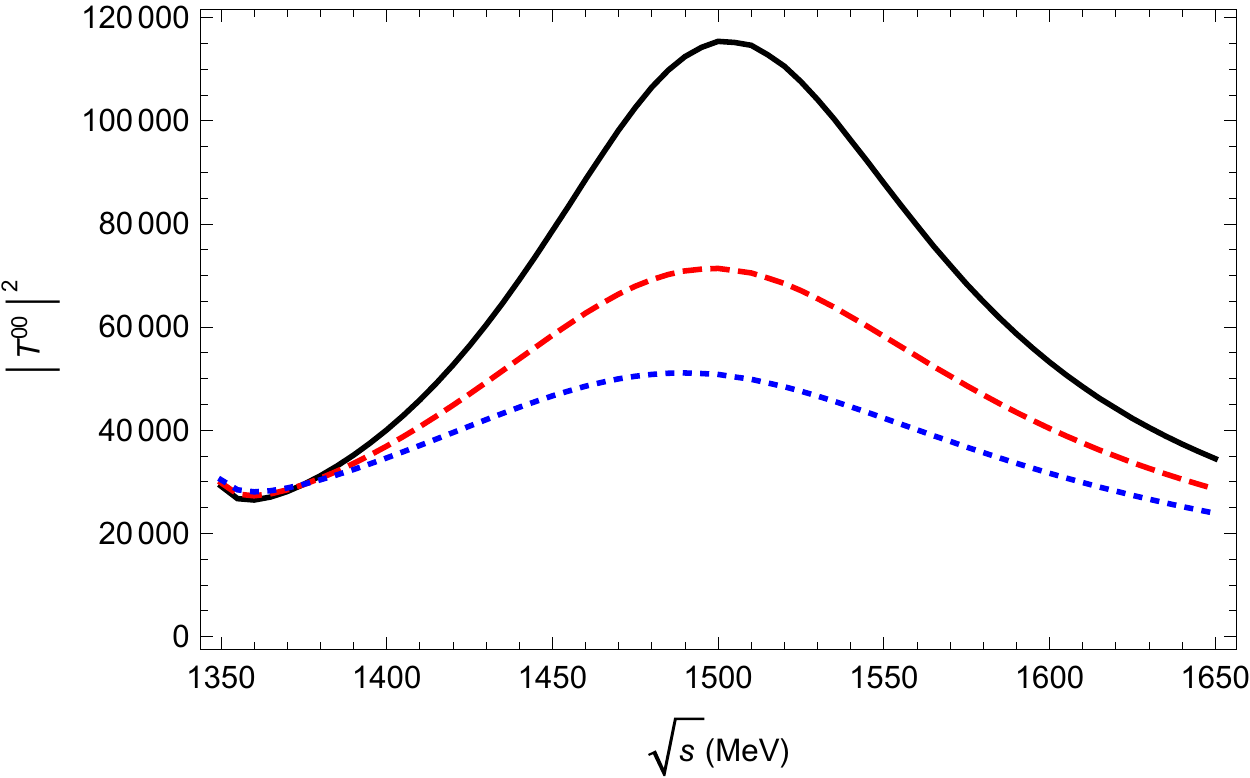} &
 \includegraphics[width=0.48\textwidth]{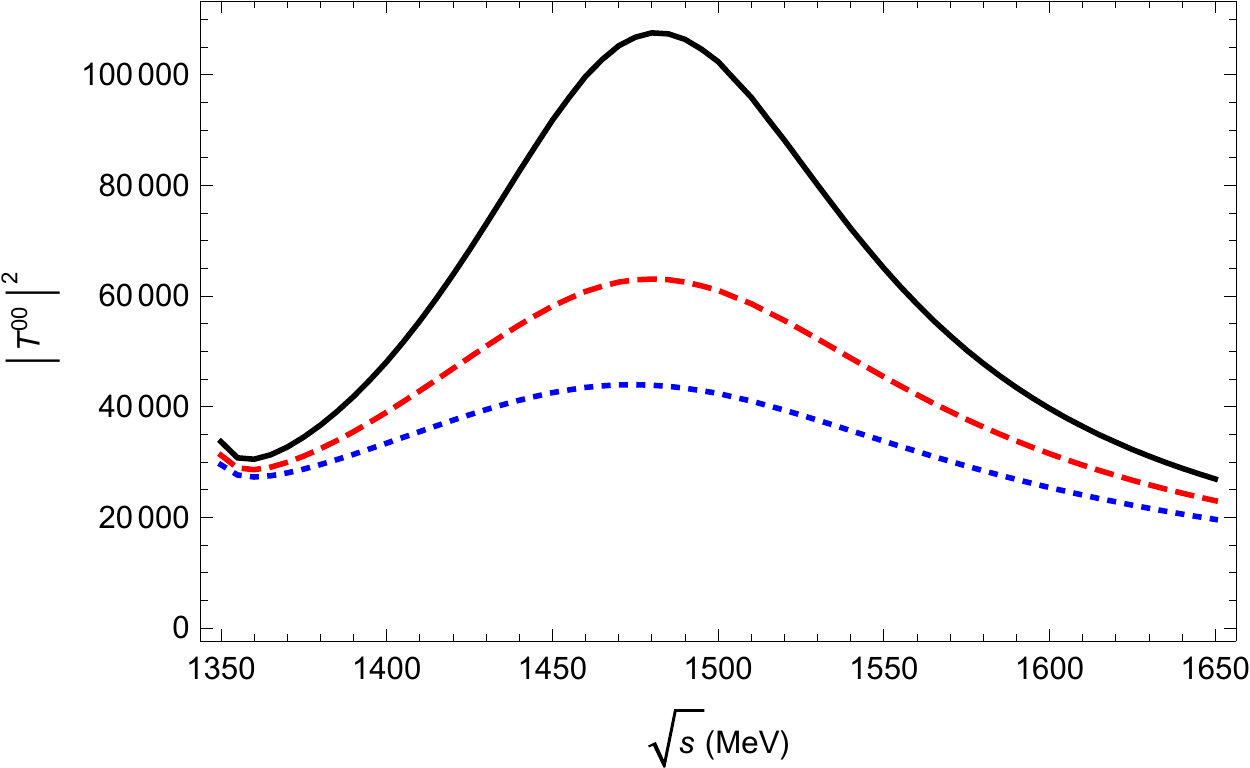}
\end{tabular}
\caption{$|T^{(00)}(s)|^2$ with the $\pi\pi$ box diagram contribution included for 
different values of $\Lambda$ in $F(q)$, 
cf. Eq.~\eqref{111016.2}. 
Specifically, $\Lambda=1200$ (black solid line), 1300 (red dashed line) 
and 1400~MeV (blue dotted line), for $q_{\rm max}=875$ (left panel) 
and $q_{\rm max}=1000$~MeV (right panel).}
 \label{fig:box}
 \end{figure}
\end{center}

The inclusion of the $\pi\pi$ box diagram, on top of the convolution with the 
$\rho$ mass squared distribution for calculating the $g(s)$ function,
does not alter the previous conclusion on the absence of a pole in the 
isoscalar tensor channel. 
However, the isoscalar scalar pole  develops a larger width around 200-300~MeV, that increases with $\Lambda$, as 
can be inferred from Fig.~\ref{fig:box}, 
where we plot $|T^{(00)}(s)|^2$. On the other hand, the position of the peak barely changes compared to the one 
given in the last two rows of Table~\ref{Polegc}.
 Tentatively this pole could be associated to 
the $f_0(1370)$ resonance, which according to Refs.~\cite{albaladejo.101016.1,bugg.101016.2} 
decays mostly to $\pi\pi$ with a width around 200~MeV. In the PDG \cite{pdg.071016.3} 
the total width of the $f_0(1370)$ is given with a large uncertainty, 
within the range 200-500~MeV and the $\pi\pi$ decay mode is qualified as dominant. 
 The nearby $f_0(1500)$ resonance has a much smaller width, around 100~MeV, and 
its coupling and decay to $\pi\pi$ 
 is suppressed. These properties of the $f_0(1500)$ are discussed in detail in 
Ref.~\cite{albaladejo.101016.1}.

\subsection{Coupled-channel scattering}
\label{css}

We now consider the impact on our results when allowing for the coupling between 
channels with different orbital angular momenta, an issue not considered in 
Ref.~\cite{oset.081016.1}.   In Table~\ref{coupledch} we show the different 
channels that couple for given $JI$ quantum numbers and pay special attention 
to the $(J,I)=(0,0)$ and $(2,0)$ channels. 
 Apart from the conservation of $J$ and $I$, one also has to impose 
invariance under parity, which avoids the mixing between odd and even $\ell$'s. 

\begin{table}[t]
\begin{center}
\begin{tabular}{|c|c|} \hline
$(J,I)$ & ($\ell, S$) channels   \\ \hline \hline
$(0,0)$ & $(0,0)$, $(2,2)$\\ \hline
$(2,0)$ & $(0,2)$, $(2,0)$, $(2,2)$ \\ \hline
\end{tabular}
\caption{Coupled-channels with different orbital angular momentum.}
\label{coupledch}
\end{center}
\end{table}

\begin{table}[t]
\begin{center}
\begin{tabular}{|c|c|c|c|c|c|}
\hline
$q_{\rm max}$ (MeV) & Mass (MeV) & $(\gamma^{(00)}_{00})^2$ (GeV$^2$) 
& $(\gamma^{(00)}_{22})^2$ (GeV$^2$) & $X^{(00)}_{00}$  & $X^{(00)}_{22}$  \\
\hline
775   & 1515.3 & $57.2$   & $0.2$ & $0.75$ & $0.00$ \\ \hline
         & 1386.6 & $-7.6$ & $-20.2$ &  $<0$       & $<0$   \\ \hline
875   & 1492.4 & $72 .1$  & $1.0$ & $0.69$ & $0.01$ \\ \hline
         & 1396.8 & $-13.5$ & $-19.4$ & $<0$ & $<0$  \\ \hline
1000 & 1455.3 & $116.2$ & $8.3$ & $0.80$ & $0.06$ \\ \hline
         & 1415.7 & $-53.1$ & $-25.1$ & $ <0$ & $<0$\\ \hline
\end{tabular}
\caption{Bound state poles in the partial wave amplitudes of quantum numbers $(J,I)=(0,0)$ 
with varying cutoff $q_{\rm max}$, which is indicated in the first column. The masses 
(2nd column), the residues to $(\ell, S)=(0,0)$ and $(2,2)$ (3rd and 4th columns) and 
the compositeness coefficients $X^{00)}_{00}$ and $X^{00)}_{22}$ (6th and 7th columns) 
are also given. For the lighter poles the compositeness coefficients are small and 
negative, so that they cannot be interpreted as physical states 
 contrary to common wisdom \cite{hyodo.101016.5,aceti.101016.6,sekihara.101016.7}.
\label{poles00cc}} 

\end{center}
\end{table}

When including coupled channel effects, one finds two poles in the  
channels with $(J,I)=(0,0)$, that are reported in Table~\ref{poles00cc} for 
various values of  $q_{\rm max}$ (shown in the first column). We give from left 
to right the pole mass (second column), the residues (third and fourth ones) 
and compositeness  coefficients (fifth and sixth ones) of the different channels, 
$(\ell, S)=(0,0)$ and $(2,2)$, respectively. 
One of the poles is heavier and closer to the $\rho\rho$ threshold with similar properties  
as the pole in the uncoupled case, compare with Table~\ref{Polegc}, particularly for 
$q_{\rm max}=775$~MeV.  Nonetheless, as $q_{\rm max}$ increases 
the difference of the properties of this pole between the coupled and uncoupled cases 
is more pronounced.  In particular let us remark that now $X_{00}^{(00)}$  is 
always $\gtrsim 0.7$ and for $q_{\rm max}=1$~GeV the residue to the channel $(\ell, S)
=(0,0)$ is much larger than in the uncoupled case. 
 Additionally, we find now a lighter  pole which lays above the branch point 
singularity at $\sqrt{3}\,m_\rho\simeq 1343$~MeV. 
For lower values of the cutoff $q_{\rm max}$ this pole couples more strongly to the 
$(\ell, S)=(2,2)$ channel, but as the cutoff increases its residue for the
channel $(\ell, S)=(0,0)$  also increases in absolute value 
  and it is the largest for $q_{\rm max}=1$~GeV. 
It is then clear that  both channels $(\ell, S)=(0,0)$ and $(2,2)$ are relevant for 
the origin of  this pole. 
Note that the residues for this lighter pole are negative, which is at odds with 
the standard interpretation of the residue $(\gamma^{(00)}_{\ell S})^2$ of a bound state 
as the coupling squared.
  This implies that the compositeness coefficients 
$X^{(00)}_{\ell S}$ are all negative, which is at odds with a probabilistic 
interpretation as suggested in Refs.~\cite{hyodo.101016.5,aceti.101016.6,sekihara.101016.7} 
for bound states. 
The moduli of the $|X^{(00)}_{\ell S}|$ are all small because this lighter pole lays
quite far from the $\rho\rho$ threshold. 
The fact that its mass is not far from the strong branch point singularity at 
$\sqrt{3}\,m_\rho$ makes that this pole is very much affected by the left-hand 
cut discontinuity. In this respect, it might well be  that the presence of 
this pole with anomalous properties  is just an artefact of the unitarization 
formula of Eq.~\eqref{081016.17}, that treats the left-hand cut discontinuity of 
the potential perturbatively.  One can answer  this question by solving exactly 
the $N/D$ method \cite{albaladejo.261016.3,guo.261016.4},
 so that the left-hand cut discontinuity of the potential is properly treated and 
the resulting amplitude has the right analytical properties. 
Let us recall that Eq.~\eqref{081016.17} is an approximate algebraic solution of 
the $N/D$ method by treating perturbatively the left-hand cut discontinuities of 
the coupled partial waves \cite{oller.091016.12,oller.261016.2,meissner.091016.9}. 
For the uncoupled scattering such effects are further studied in detail in the next section.

\begin{table}[htb]
\begin{center}
\begin{tabular}{|c|c|c|c|c|}
\hline
$q_{\rm max}$ (MeV) & Mass (MeV) & $(\gamma^{(20)}_{02})^2$ (GeV$^2$) & $(\gamma^{(20)}_{20})^2$ (GeV$^2$) &  $(\gamma^{(20)}_{22})^2$ (GeV$^2$)  \\
\hline
775   & 1355.1 & $-0.8$   & $-0.0$ & $-8.6$\\ \hline
875   & 1358.2 & $-0.7$  & $-0.0$ & $-8.3$  \\ \hline
1000 &  1361.8 & $-0.6$ & $-0.0$ & $-7.9$  \\ \hline
\end{tabular}
\caption{Bound state poles in the partial wave amplitudes of quantum numbers $(J,I)=(2,0)$ with  different cutoffs $q_{\rm max}$, indicated in the first column. The masses (2nd column) and the residues to $(\ell, S)=(0,2)$, $(2,0)$ and $(2,2)$ (3rd, 4th and 5th columns) 
 are shown. Here, ``$-0.0$'' denotes a small but negative number.} 
\label{poles20cc}
\end{center}
\end{table}

For the $(J,I)=(2,0)$ partial waves we have three coupled channels, 
$(\ell, S)=(0,2)$, $(2,0)$ and $(2,2)$ and, contrary to the uncoupled case, we 
now find a pole that  lays above the branch point singularity. We give its mass and 
residues for different $q_{\rm max}$ in Table~\ref{poles20cc}, with the same notation as in 
Table~\ref{poles00cc}. Notice that these pole properties  are very stable under the 
variation of $q_{\rm max}$. This pole couples by far much more 
strongly to the channel with $(\ell, S)=(2,2)$ than to any other channel. This 
indicates that it is mainly due to the dynamics associated with the $(\ell, S)=(2,2)$ 
channel. But  the same comments are in order here as given above for the lighter 
isoscalar scalar pole, because its residues shown in Table~\ref{poles20cc} are negative 
and so are the corresponding compositeness coefficients. Hence, the lighter pole for 
$(J,I)=(0,0)$ and the one found for $(2,2)$ cannot be considered as robust 
results of our analysis.  This has to be contrasted to the case of the heavier 
isoscalar scalar  pole that is stable under relativistic corrections, coupled-channel effects 
and has quite standard properties regarding its couplings and compositeness coefficients.

\section{First iterated solution of the $N/D$ method}
\label{nd.211216.1}

In this section for definiteness we only consider uncoupled scattering. We have in mind the $(J,I)=(0,0)$ and $(J,I)=(2,0)$ 
quantum numbers to which  special attention has been paid in the literature concerning the generation of poles 
that could be associated with the $f_0(1370)$ and $f_2(1270)$ resonances, as discussed above. Further applications 
of the improved unitarization formalism presented in this section are left for future work.
 
According to the $N/D$ method \cite{chew} a partial-wave amplitude can be written as
\begin{align}
\label{181216.1}
T&=\frac{N(s)}{D(s)}~,
\end{align}
where the function $D(s)$ has only the unitarity or right-hand cut (RHC) while $N(s)$ only has 
the left-hand cut (LHC).  
The secular equation for obtaining  resonances and bound states corresponds to look for the zeros of $D(s)$, 
\begin{align}
\label{181216.2}
D(s_i)=0~.
\end{align}
Below threshold along the real $s$ axis this equation is purely real because $D(s)$ has 
a non-vanishing imaginary part only for $s>s_{\rm th}$.

 However, with our unitarization procedure from leading-order unitary chiral perturbation theory (UChPT) 
 we have obtained the approximation 
\begin{align}
\label{181216.3}
T^{(JI)}(s)&=\frac{V^{(JI)}(s)}{1-V^{(JI)}(s)g_c(s)}~,
\end{align}
and the resulting equation to look for the bound states is
\begin{align}
\label{181216.4}
D_U(s)&=1-V(s)g_c(s)=0~.
\end{align}
 Notice that Eq.~\eqref{181216.3}, contrary to the general  Eq.~\eqref{181216.2}, 
has an imaginary part below the branch-point singularity at $s=3m^2$. 

We can go beyond this undesired situation by considering the first-iterated solution to the $N/D$ method. This is indeed
similar to  Eq.~\eqref{181216.3} but improving upon it because it allows us to go beyond the 
on-shell factorization employed in this equation. In the first-iterated $N/D$ solution one identifies 
the numerator function $N(s)$ to the tree-level calculation $V^{(JI)}(s)$ and employs the exact dispersive expression 
for $D(s)$. Namely, it reads\footnote{A comprehensive introduction to the $N/D$ method is given in 
Refs.~\cite{guo.261016.4,oller.091016.12,oller.221216.1,martin.290916.1}.}
\begin{align}
\label{181216.5}
N(s)&=V^{(JI)}(s)~,\nn\\
D(s)&=\gamma_0+\gamma_1(s-s_{th})+\frac{1}{2}\gamma_2(s-s_{\sth})^2
+\frac{(s-\sth)s^2}{\pi}\int_{\sth}^\infty ds'\frac{\rho(s')V^{(JI)}(s')}{(s'-\sth)(s'-s)(s')^2}~,\nn\\
T_{ND}(s)&=\frac{N(s)}{D(s)}~,
\end{align}
 with the phase space factor $\rho(s)$ given by $\rho(s)=\sigma(s)/16\pi$.
We have taken three subtractions in the dispersion relation for $D(s)$ because $V^{(JI)}(s)$ diverges as $s^2$ for $s\to \infty$.  

From our present study  we have concluded that $T^{(JI)}(s)$ is stable in the threshold region 
under relativistic corrections as well as under the addition of coupled channels. 
 Because of the stability of the results in this region under relativistic corrections and  
by visual inspection of the potentials in Fig.~\ref{fig:potentials} one concludes that  the near-threshold region 
is quite safe of the problem related to the branch-point singularity of the LHC associated 
with one-$\rho$ crossed-channel exchanges.
 We then determine the subtractions constants, $\gamma_0$, $\gamma_1$ and $\gamma_2$ 
 in $D(s)$ by matching $T_{ND}(s)$, Eq.~\eqref{181216.5}, and $T^{(JI)}(s)$, Eq.~\eqref{181216.3}, around threshold 
($\sth$).
 At the practical level it is more convenient to match 
$1/T(s)$, so that in the threshold region up to ${\cal O}(s^3)$ one has:
\begin{align}
\label{181216.6}
\gamma_0+\gamma_1 (s-s_{th})+\frac{1}{2}\gamma_2(s-s_{\sth})^2
&=1-V(s)g_c(s)-\frac{(s-\sth)s^2}{\pi}\int_{\sth}^\infty ds'\frac{\rho(s')V(s')}{(s'-\sth)(s'-s)(s')^2}
\nn\\
&\equiv  \omega(s)~.
\end{align}
 In this way, 
\begin{align}
\label{181216.7}
\gamma_0=&1-V(\sth)g_c(\sth)~,\nn\\
\gamma_1=&\omega'(\sth)~,\nn\\
\gamma_2=&\omega''(\sth)~.
\end{align}

 The dependence of our present results on the cutoff used in $T^{(JI)}(s)$ 
stems from the matching conditions of Eq.~\eqref{181216.7}.
 However, let us stress that the analytical properties of $D(s)$ and $N(s)$ are correct, 
they have the RHC and LHC with the appropriate extent and branch point singularities, respectively, 
and the resulting amplitude is unitarized.

 \begin{figure}[t]
 \begin{center}
 \includegraphics[width=0.4\textwidth,angle=-90]{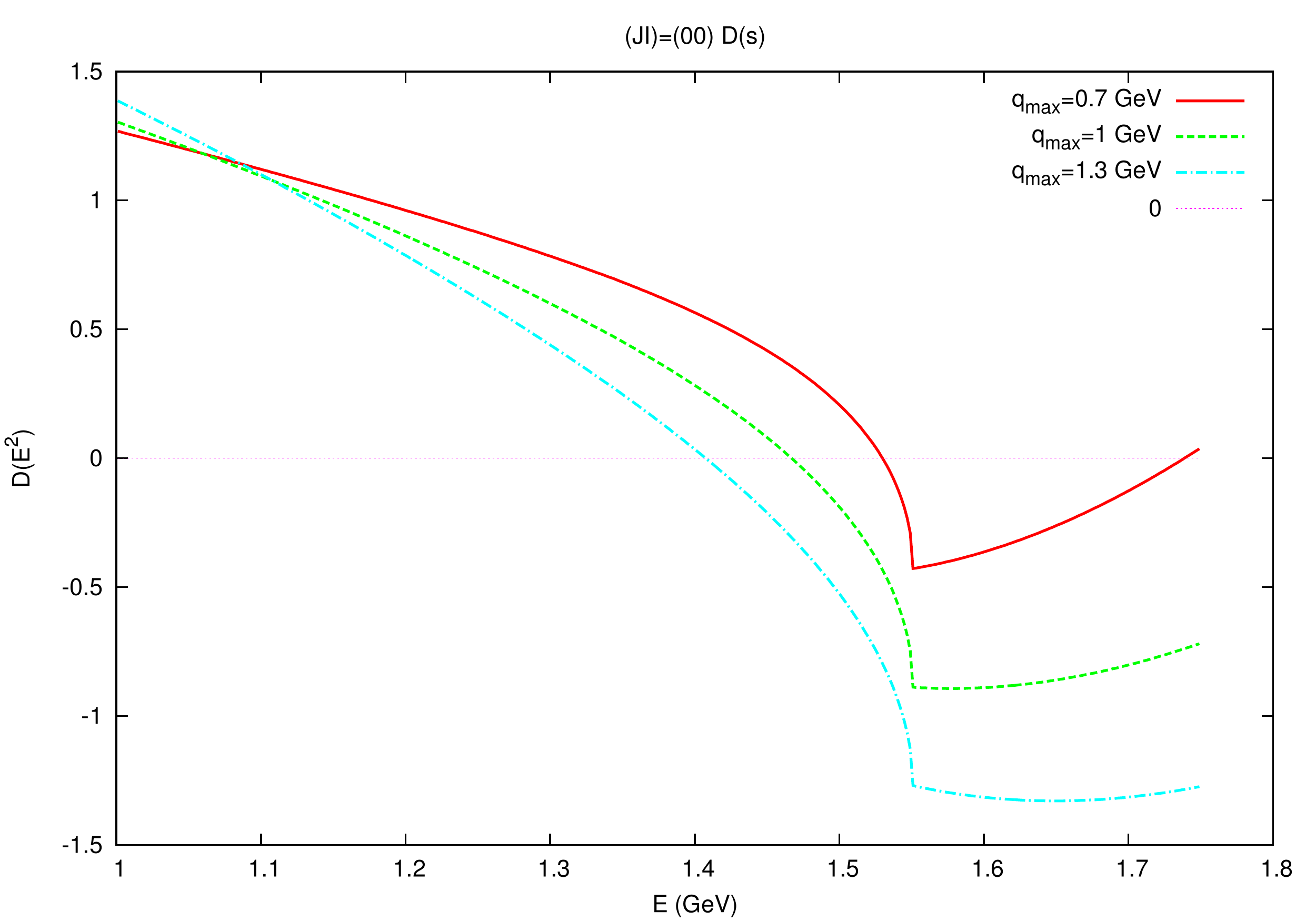} 
\caption{$D(s)$ function, Eq.~\eqref{181216.5}, for $(J,I)=(0,0)$. 
Above threshold only the real part is shown.
  \label{fig.181216.4}}
\end{center}
 \end{figure}

\subsection{Results $J=0$, $I=0$}
\label{sec.181216.3}

\begin{figure}[ht]
 \begin{center}
\begin{tabular}{cc}
 \includegraphics[width=0.34\textwidth,angle=-90]{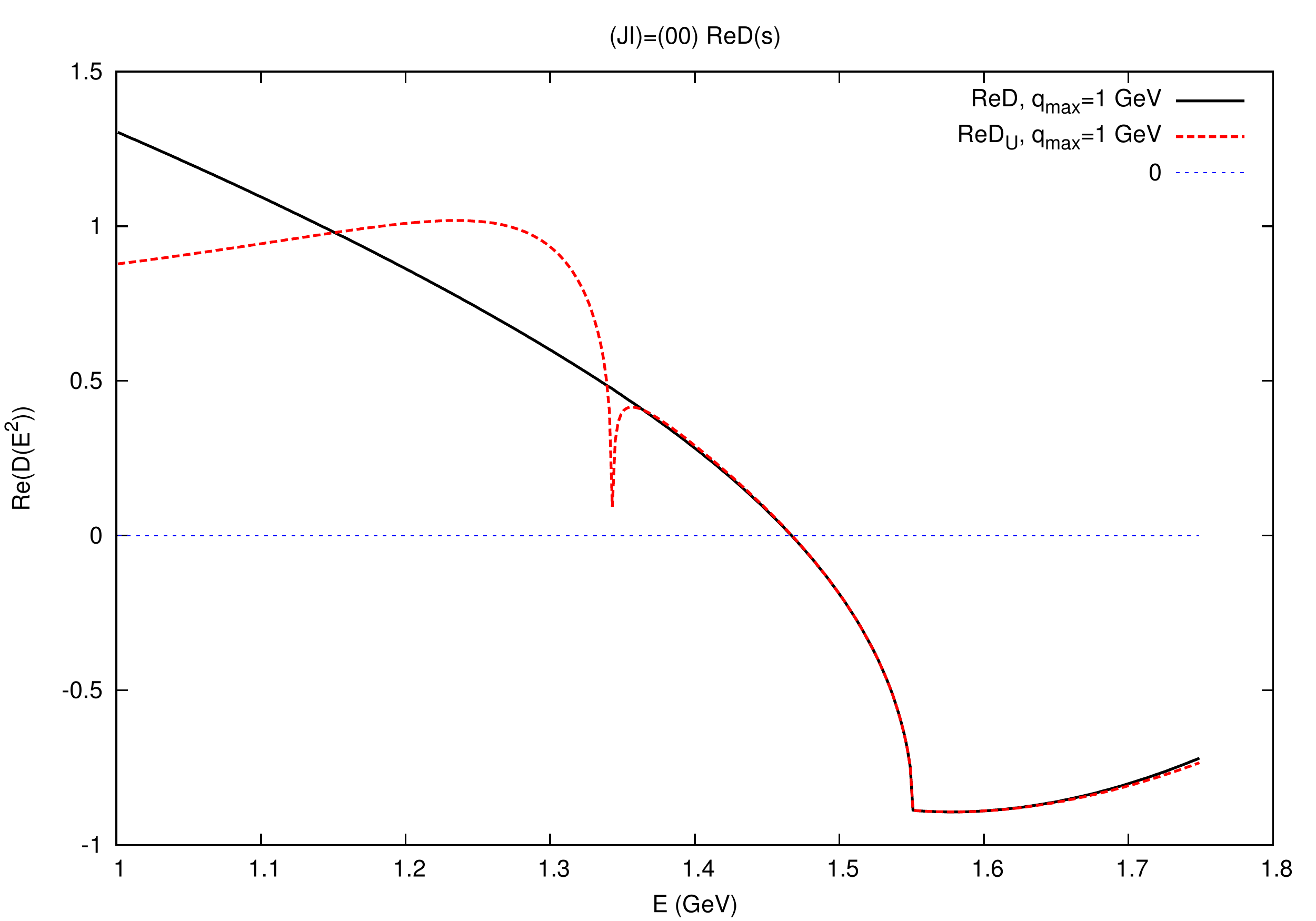} &
 \includegraphics[width=0.34\textwidth,angle=-90]{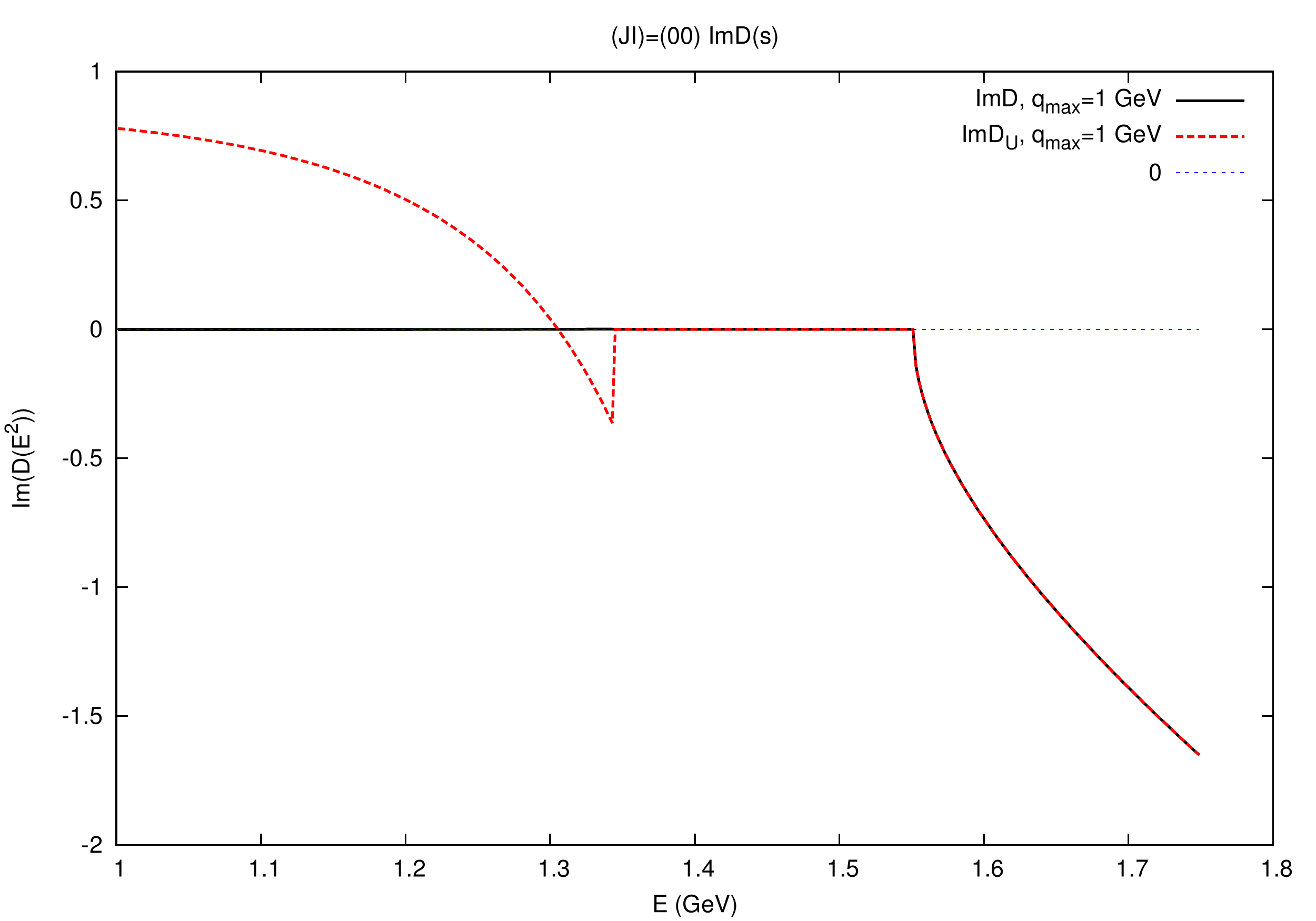}
\end{tabular}
\caption{$(J,I)=(0,0)$. Real (left) and imaginary (right panel) parts of $D(s)$ compared also with $D_U(s)=1-V^{(00)}(s)g_c(s)$ from 
leading-order UChPT. 
 \label{fig.181216.5}}
\end{center}
 \end{figure}

We plot $D(s)$ for $(J,I)=(0,0)$ in Fig.~\ref{fig.181216.4} for $q_{\rm{max}}=0.7$, 1 and 1.3~GeV by the red solid, 
green dashed and blue dash-dotted lines, in order. 
The crossing with the zero line (dotted one) indicates the mass of the bound state. 
This mass decreases with increasing $q_{\rm max}$, 
being around 1.4~GeV for the largest cutoff and very close to threshold for the smallest.
In Fig.~\ref{fig.181216.5} we compare the real (left) and imaginary parts (right panel) of 
$D(s)$ and $D_U(s)=1-V(s)g_c(s)$ for a cutoff of 1~GeV. We do not show more values of the cutoff because the 
same behavior results. 
The function $D(s)$ and $D_U(s)$ match up to around the branch-point singularity at $\sqrt{s}=3m_\rho^2$. 
Below it $D_U(s)$ becomes imaginary, cf. Eq.~\eqref{181216.4}, while  
 $D(s)$ remains real and has this right property by construction, cf. Eq.~\eqref{181216.5}. 
Above threshold the imaginary parts of both functions coincide as demanded by unitarity. 
We see that for these quantum numbers our new improved unitarization formalism and the one used to derive 
Eq.~\eqref{081016.17} agree very well. 
  The bound-state mass 
remains  the same as given in Table~\ref{Polegc} because the functions $D(s)$ and $D_U(s)$ match perfectly 
well in the region where these poles occur, as it is clear from  Fig.~\ref{fig.181216.5}. 
 This should be  expected because for $(J,I)=(0,0)$ the branch 
point singularity was much weaker than for other cases, e.g. $(J,I)=(2,0)$, as discussed above.

 \begin{figure}[ht]
 \begin{center}
 \includegraphics[width=0.4\textwidth,angle=-90]{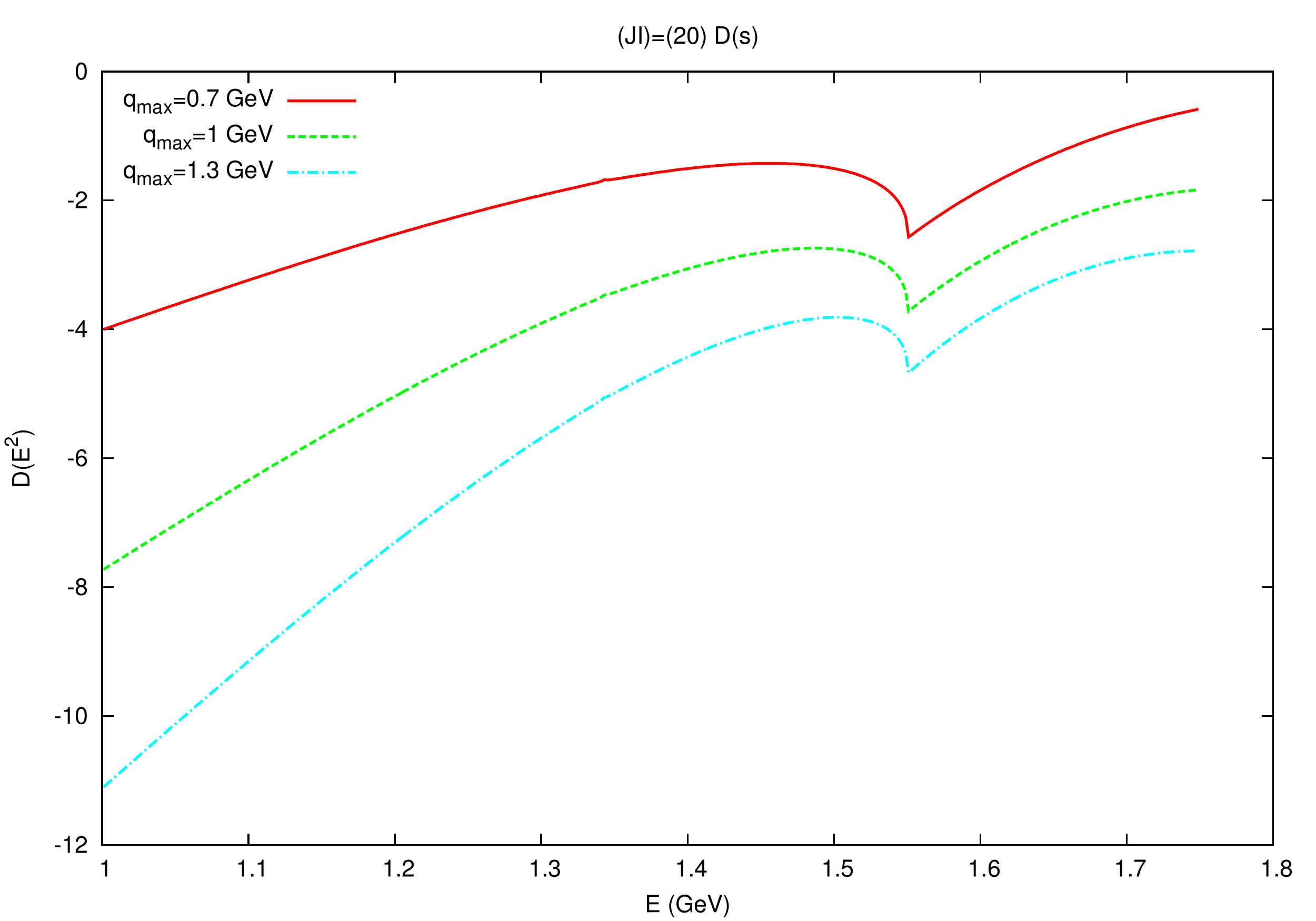} 
\caption{$D(s)$ function, Eq.~\eqref{181216.5}, for $(J,I)=(2,0)$. 
Above threshold only the real part is shown.
  \label{fig.181216.1}}
\end{center}
 \end{figure}

\subsection{Results $J=2$, $I=0$}
\label{sec.181216.2}

\begin{figure}[htb]
 \begin{center}
\begin{tabular}{cc}
 \includegraphics[width=0.34\textwidth,angle=-90]{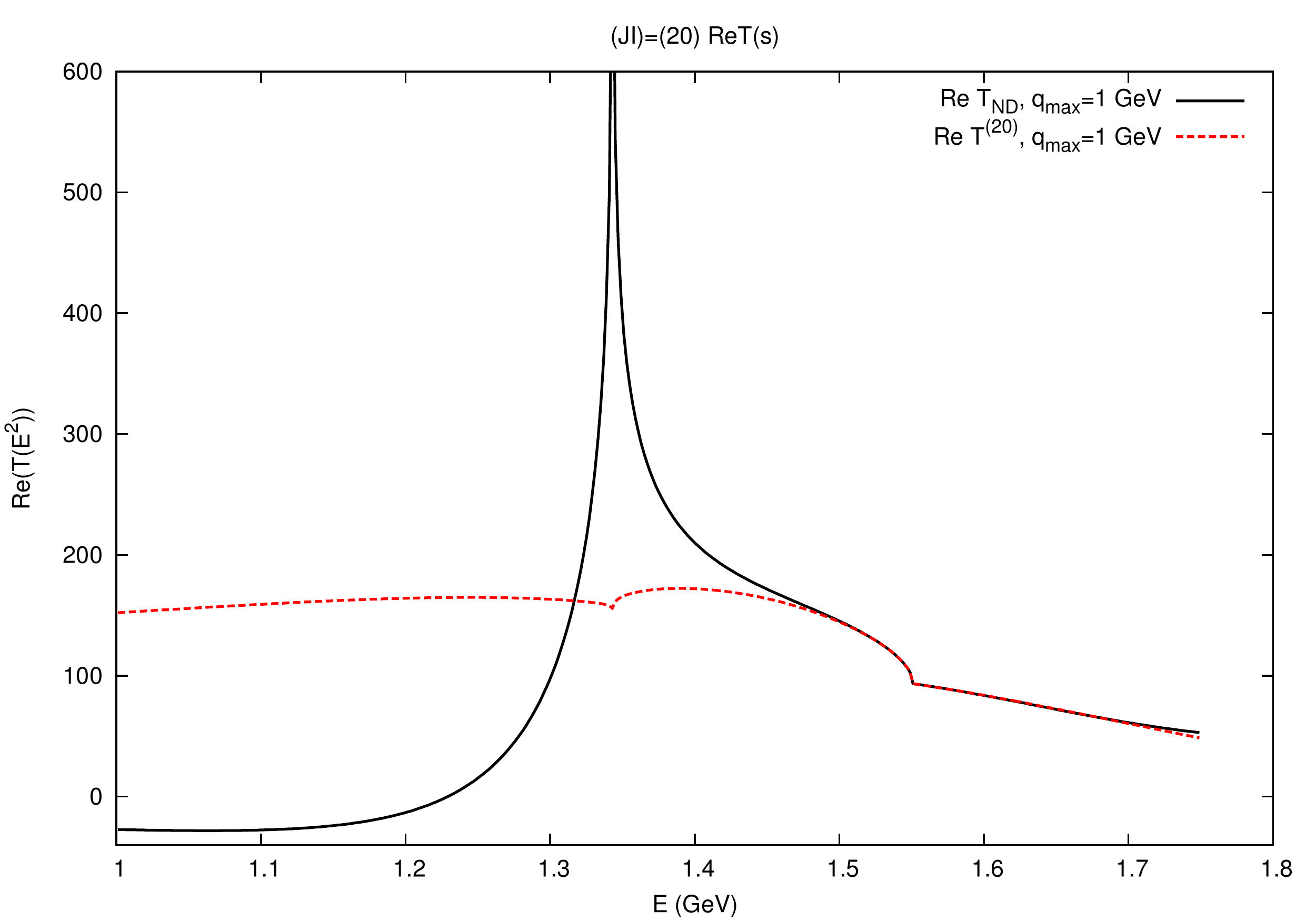} &
 \includegraphics[width=0.34\textwidth,angle=-90]{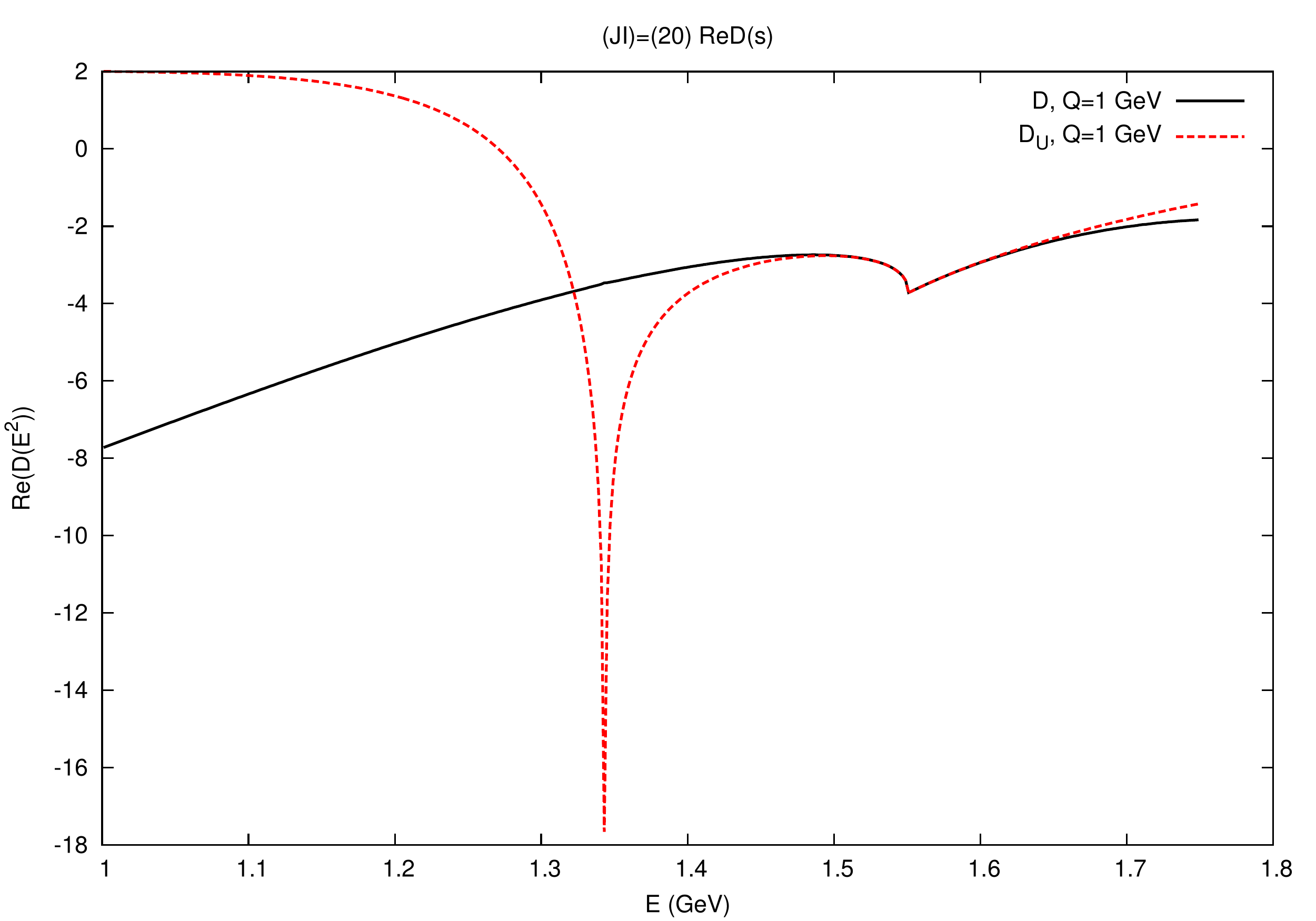} \\
\end{tabular}
\caption{$(J,I)=(2,0)$. Left panel: real part of $T_{ND}(s)$ (black solid) compared with that of $T^{(20)}(s)$ (red dashed). 
 Right panel: comparison between the real parts of $D(s)$ (black solid) and $D_U(s)=1-V^{(20)}(s)g_c(s)$ (red dashed).
 \label{fig.181216.2}}
\end{center}
 \end{figure}

We plot $D(s)$ for $(J,I)=(2,0)$ in Fig.~\ref{fig.181216.1} for $q_{\rm{max}}=0.7$, 1 and 1.3~GeV by the red solid, green dashed 
and blue dash-dotted lines, respectively. One can see that in this region the function $D(s)$ is large and negative so that 
there is by far no pole in the $f_2(1270)$ region.  
In order to show the curves more clearly we use only $q_{\rm max}=1$~GeV in Fig.~\ref{fig.181216.2}, for other values of 
$q_{{\rm max}}$ the behavior is the same. In the left panel we compare the real parts  of 
$T_{ND}(s)$ (black solid) and $T^{(20)}(s)$ (red dashed) while in the right panel we proceed similarly for the 
real parts of $D(s)$ and $D_U(s)=1-V^{(20)}(s) g_c(s)$, with the same type of lines in order. 
 All the functions match near the threshold region and above it, but they strongly depart once we approach the LHC 
branch-point singularity at $s=3m_\rho^2$ and beyond (for smaller values of $s$). 
 Notice that $D(s)$, which has not such branch-point singularity, follows 
then the smooth decreasing trend already originated  for $3m^2<s<4m^2$. For $(J,I)=(0,0)$ the 
corresponding smooth trend is that of a decreasing function, cf. Fig.~\ref{fig.181216.4}. 
The branch-point singularity is clearly seen in $T^{(20)}(s)$ because it is proportional to 
$V^{(20)}(s)$.

In summary, the conclusions obtained in Sec.~\ref{uss} regarding the generation of the pole that 
could be identified with the $f_0(1370)$ and the absence  of that associated with the $f_2(1270)$ as claimed in Ref.~\cite{oset.081016.1}
fully hold. As a matter of fact, they get reinforced after considering the more elaborated unitarization process that is obtained
 here by taking the first-iterated $N/D$ solution.

\section{Summary and conclusions}
\label{sec:summ}

In this paper, we have revisited the issue of resonance generation in unitarized
$\rho\rho$ scattering using a chiral covariant formalism. The main results of
our study can be summarized as follows:
\begin{itemize}
\item[i)] We have developed a partial-wave projection formalism that is applicable
 to  the covariant treatment  of $\rho\rho$ scattering.  
 In particular, we point out that accounting for the  full $\rho$-meson propagator leads to a 
branch point in the partial wave projected amplitudes at $s = 3m_\rho^2 \simeq 1.8\,$GeV$^2$,
about 208~MeV below the $2\rho$ threshold.  This branch point does not appear in the
extreme non-relativistic treatment of the propagator.
\item[ii)] Evaluating the $T$-matrix using the standard form, see Eq.~(\ref{081016.17}) that 
treats the left-cut perturbatively, we find a pole in the scalar isoscalar channel
close to the $\rho\rho$ threshold that can be associated with the $f_0(1370)$ 
resonance, in agreement with the findings of Ref.~\cite{oset.081016.1}, though there
are minor quantitative differences.
\item[iii)] In contrast to  Ref.~\cite{oset.081016.1}, we do not find a tensor state below the scalar one.
This can be traced back to the influence of the aforementioned branch point. We therefore conclude
that the state that is identified in Ref.~\cite{oset.081016.1} with the $f_2(1270)$ is an artifact of the
non-relativistic approximation  and its generation does not hold from the arguments given in that reference. 
 \item[iv)] We have also worked out the effects of the coupling between channels with different orbital 
 angular momenta,  which lead to additional states. These, however, have negative composite 
 coefficients and are thus not amenable to a simple bound state interpretation. As these states are
 close to the branch point at   $s = 3m_\rho^2$, the perturbative treatment of the left-hand cut,
 as employed here, 
 is certainly not sufficient to decide about their relevance. 
\item[v)] We have improved the treatment of the left-hand cut by employing the first-iterated $N/D$ method, 
in particular this method avoids the factorization approach of leading order UChPT. We worked the solutions 
that follow for uncoupled scattering in the $(J,I)=(0,0)$ and $(2,0)$ channels. The outcome fully agrees with 
the conclusions already obtained from UChPT and, notably, the absence of a pole that could be associated with the $f_2(1270)$ 
is firmly reinforced. 

\end{itemize} 

A lesson from points iii) and v) is clear. A strongly attractive interaction in a given channel is a necessary but by 
far not sufficient condition to generate a multi-hadron bound state. This argument, as used in Ref.~\cite{oset.081016.1},
 is in general terms too naive because it does not take into account the possible raise of a singularity in the true potential 
 between the range of validity of the approximation used and the predicted bound-state mass from the latter. 
It could be rephrased as trying to deduce the values of the function $1/(1+x)$ for $x<-1$ by knowing its values for $x$ 
around $0$.

We conclude that the approach presented here should be used to investigate the possible generation
of meson resonances from the interaction of vector mesons. In the next steps, we will investigate how 
the relativistic effects affect the conclusions of the SU(3) calculation of Ref.~\cite{Geng:2008gx} and
will further sharpen the framework along the lines mentioned, in particular by solving exactly 
 the $N/D$ equations \cite{guo.261016.4}.

\section*{Acknowledgements}

We thank Maxim Mai for useful discussions and for his contribution during the early stages of this 
investigation. We would also like to thank E. Oset for some criticism which led us to
 add some additional material to the manuscript. 
This work is supported in part by the DFG (SFB/TR~110, ``Symmetries and the
Emergence of Structure in QCD'').
JAO would like to acknowledge partial financial support from 
 the MINECO (Spain) and ERDF (European Commission) grant FPA2013-40483-P and
 by the Spanish Excellence Network on Hadronic Physics FIS2014-57026-REDT. 
The work of UGM was supported in part by The Chinese Academy of Sciences 
(CAS) President's International Fellowship Initiative (PIFI) grant no. 2015VMA076.

\appendix

\section{Partial-wave projection formalism}
\label{app.081016.1}
\def\theequation{\Alph{section}.\arabic{equation}}
\setcounter{equation}{0}

In this appendix we detail the projection formalism used in this work to calculate the 
different $\rho\rho$ partial waves. 
First, we give the expression for the polarization vectors for a massive spin-one particle 
with a three-momentum $\vp$ and third component of spin $\sigma$ 
in the $z$ axis of its rest frame, that we denote by $\ve(\vp,\sigma)$. In the 
rest frame they are given by
\begin{align}
\label{200916.1}
\ve(\mathbf{0},\sigma)=&\left(
\begin{array}{c}
0\\
\boldsymbol\varepsilon_\sigma
\end{array}
\right)~,
\end{align}
with
\begin{align}
\label{260916.2}
\boldsymbol\ve_0=\left(
\begin{array}{c}
0\\
0\\
1
\end{array}\right)~,~
\boldsymbol\ve_{\pm 1}=\frac{\mp 1}{\sqrt{2}} \left(
\begin{array}{c}
1\\
\pm i\\
0
\end{array}\right)~.
\end{align}
Next, we take a Lorentz transformation $U(\vp)$ along the vector $\vp$ 
 that takes the particle four-momentum at rest to its final value,
\begin{align}
\label{230916.1}
U( \vp)\left(
\begin{array}{c}
m\\
\mathbf{0}
\end{array}
\right)=&
\left(
\begin{array}{c}
E_p\\
\vp
\end{array}
\right)~,
\end{align}
with $E_p=\sqrt{m^2+\vp^2}$. 
We also introduce the rotation $R(\up)$ that takes $\hat{\mathbf{z}}$ to $\up$,
\begin{align}
\label{230916.2}
R(\up)\hat{\mathbf{z}}=&\up~.
\end{align}
In terms of the polar ($\theta$) and azimuthal ($\phi$) angles of $\up$ this rotation is defined as
\begin{align}
\label{230916.3}
R(\up)=&R_z(\phi)R_y(\theta)~,
\end{align}
with the subscripts $z$ and $y$ indicating the axis of rotation. 
For latter convenience we write the Lorentz transformation $U(\vp)$ as
\begin{align}
\label{230916.4}
U(\vp)=&R(\hat{\vp})B_z(|\vp|)R(\up)^{-1}~,
\end{align}
where $B_z(|\vp|)$ is a boost along the $\hat{\mathbf{z}}$ axis with velocity $v=-\beta$ and 
$\beta=|\vp|/E_p$. 
 Namely,
\begin{align}
\label{230916.6}
B_z(|\vp|)= \left(
\begin{array}{cccc}
\gamma & 0 & 0 & \gamma\beta \\
0           & 0 & 0 & 0 \\
0           & 0 & 0 & 0 \\
\gamma \beta & 0 & 0 & \gamma
\end{array}
\right)
\end{align}
and
\begin{align}
\label{230916.7}
\gamma=\frac{1}{\sqrt{1-\beta^2}}~.
\end{align}
Notice that one could also include any arbitrary rotation around the $\hat{\mathbf{z}}$ axis to the right end of 
 Eq.~\eqref{230916.3}.  Of course, this does not have any affect on either Eqs.~\eqref{230916.2} and  \eqref{230916.4}
(for the latter one let us note that $B_z(|\vp|)$ commutes with a rotation around the $\hat{\mathbf{z}}$ axis).

The action of $U(\vp)$ on $\epsilon(\mathbf{0},\sigma_i)$ gives us the polarization vectors with definite three-momentum $\vp$,
 whose expressions are 
\begin{align}
\label{260916.1}
\ve(\mathbf{p},0)=\left(
\begin{array}{c}
\gamma\beta\cos\theta\\
\frac{1}{2}(\gamma-1)\sin2\theta\cos\phi\\
\frac{1}{2}(\gamma-1)\sin2\theta\sin\phi\\
\frac{1}{2}\left(1+\gamma+(\gamma-1)\cos2\theta\right)
\end{array}\right)~,~
\ve(\mathbf{p},\pm 1)=\mp \frac{1}{\sqrt{2}}\left(
\begin{array}{c}
\gamma\beta e^{\pm i\phi}\sin\theta\\
1+(\gamma-1)e^{\pm i\phi}\sin^2\theta\cos\phi\\
\pm i+(\gamma-1)e^{\pm i \phi}\sin^2\theta \sin\phi\\
\frac{1}{2}(\gamma-1)e^{\pm i\phi}\sin2\theta
\end{array}\right)~.
\end{align}
 The previous equation can be written in more compact form as 
\begin{align}
\label{260916.3}
\ve(\vp,\sigma)=\left(
\begin{array}{c}
\gamma\beta \hat{\vp}\cdot \boldsymbol\ve_\sigma \\
\boldsymbol\ve_\sigma +  \hat{\vp}\,(\gamma-1) \hat{\vp} \cdot \boldsymbol \ve_\sigma
\end{array}
\right)~.
\end{align}

In terms of the polarization vectors in Eq.~\eqref{260916.3} we can write the vector field 
for the neutral $\rho^0$ particle, $\rho^0_\mu(x)$, as
\begin{align}
\label{280916.1}
\rho^0_\mu(x)=\sum_\sigma \int\frac{d^3p}{(2\pi)^32E_p}\left\{
\varepsilon(\vp,\sigma)_\mu e^{-ip \,x} a(\vp,\sigma) +
 \varepsilon(\vp,\sigma)^*_\mu e^{ip \,x} a(\vp,\sigma)^\dagger
\right\}~,
\end{align}
with the corresponding similar expressions for the $\rho^\pm(x)$ fields. 
Here $a(\vp,\sigma)$ and $a(\vp,\sigma)^\dagger$ refer to the annihilation and creation 
operators, with the canonical commutation relation
\begin{align}
\label{280916.3}
[a(\vp',\sigma'),a(\vp,\sigma)^\dagger]=\delta_{\sigma\sigma'} (2\pi)^3 2E_p \delta(\vp-\vp')~.
\end{align}
In order to check the time-reversal and parity-invariance properties of the vector-vector scattering 
amplitudes worked out from the chiral Lagrangians in Eq.~\eqref{071016.1} we notice that the polarization vectors in 
 Eq.~\eqref{260916.1} satisfy the following transformation properties:
\begin{align}
\label{280916.4}
\ve(-\vp,\sigma)^*=&(-1)^\sigma (-\ve(\vp,\sigma)_0,\boldsymbol{\ve}(\vp,\sigma))~,\nn\\
\ve(-\vp,\sigma)=& (-\ve(\vp,\sigma)_0,\boldsymbol{\ve}(\vp,\sigma))~.
\end{align}

 A one-particle state  $|\vp,\sigma\rangle$  is obtained by the action of the creation operators on the vacuum state,
\begin{align}
\label{280916.2}
|\vp,\sigma \rangle = & a(\vp,\sigma)^\dagger | \mathbf{0},\sigma \rangle~.
\end{align}
From Eq.~\eqref{280916.3} it follows the following normalization for such states
\begin{align}
\label{280916.5}
\langle \vp',\sigma'|\vp,\sigma \rangle = & \delta_{\sigma'\sigma} (2\pi)^3 2E_p \delta(\vp'-\vp)~.
\end{align}
Next, we consider a two-body state characterized by the CM three-momentum $\vp$ and the third components of spin
 $\sigma_1$ and $\sigma_2$ in their respective rest frames. 
This state is denoted by $|\vp,\sigma_1\sigma_2\rangle$. Associated to this, we can 
define the two-body state with orbital angular momentum $\ell$ with its third component of orbital angular momentum 
$m$, denoted by $|\ell m,\sigma_1\sigma_2\rangle$ as
\begin{align}
\label{280916.6}
|\ell m,\sigma_1\sigma_2\rangle=&\frac{1}{\sqrt{4\pi}}\int d\up\,Y_\ell^m(\hat{\vp})|\vp,\sigma_1\sigma_2\rangle~.
\end{align} 
Let us show first that this definition is meaningful because the state $|\ell m,\sigma_1\sigma_2\rangle$ transforms under the 
rotation group as the direct product of the irreducible representations associated to the orbital angular momentum $\ell$ and 
the spins $s_1$ and $s_2$  of the two particles.

  Every single-particle state $|\vp,\sigma\rangle$ under 
the action of a rotation $R$ transforms as
\begin{align}
\label{280916.7}
R|\vp,\sigma\rangle=&R U(\vp)|\boldsymbol 0,\sigma\rangle=U(\vp')U(\vp')^{-1}R U(\vp)|\boldsymbol 0,\sigma\rangle ~,
\end{align}
and $\vp'=R\vp$.\footnote{For a general Lorentz transformation these manipulations give
 rise to the  Wigner rotation \cite{martin.290916.1}.} 
It is straightforward to show that 
\begin{align}
\label{280916.8}
R=U(\vp')^{-1}R U(\vp)~.
\end{align}
 For that we explicitly write the Lorentz transformations $U(\vp')$ and $U(\vp)$ as in Eq.~\eqref{230916.4} so that
\begin{align}
\label{280916.9}
U(\vp')^{-1}R U(\vp)=&R(\hat{\vp}')B_z(|\vp|)^{-1}R(\hat{\vp}')^{-1} R R(\hat{\vp})B_z(|\vp|)R(\hat{\vp})^{-1}~.
\end{align}
Next, the product of rotations $R(\hat{\vp}')^{-1} R R(\hat{\vp})$ is a rotation around the $z$ axis, $R_z(\gamma)$, since 
it leaves invariant  $\hat{\mathbf{z}}$. Thus,
\begin{align}
\label{280916.10}
R(\hat{\vp}')^{-1} R R(\hat{\vp})=R_z(\gamma)~,
\end{align}
or, in other terms,
\begin{align}
\label{280916.11}
R(\hat{\vp}')=& R R(\hat{\vp}) R_z(\gamma)^{-1}~,
\end{align}
Taking into account Eqs.~\eqref{280916.10} and \eqref{280916.11} in  Eq.~\eqref{280916.9} 
it follows the result in Eq.~\eqref{280916.8} because $B_z(|\vp|)$ and $R_z(\gamma)$ commute. 
Then Eq.~\eqref{280916.7} implies that 
\begin{align}
\label{280916.12}
R|\vp,\sigma\rangle=&U(\vp')R|\boldsymbol{0},\sigma\rangle=\sum_{\sigma'}D^{(s)}(R)_{\sigma'\sigma}|\vp',\sigma'\rangle~,
\end{align}
with $D^{(s)}(R)$ the rotation matrix in the irreducible representation of the rotation group with spin $s$. 

Now, we can use this result to find the action of the rotation $R$ on the state $|\vp,\sigma_1\sigma_2\rangle$ 
which is the direct product of the states $|\vp,\sigma_1\rangle$ and $|-\vp,\sigma_2\rangle$ (once the trivial 
CM movement is factorized out \cite{martin.290916.1}). In this way, 
\begin{align}
\label{280916.13}
R|\vp,\sigma_1\sigma_2\rangle=&\sum_{\sigma_1',\sigma_2'}D^{(s_1)}(R)_{\sigma_1'\sigma_1} 
D^{(s_2)}(R)_{\sigma_2'\sigma_2} |\vp',\sigma_1'\sigma_2'\rangle~.
\end{align}
 We are now ready to derive the action $R$ on $|\ell m,\sigma_1\sigma_2\rangle$,
\begin{align}
\label{280916.14}
R|\ell m,\sigma_1\sigma_2\rangle=&\sum_{\sigma_1',\sigma_2'}D^{(s_1)}(R)_{\sigma_1'\sigma_1} 
D^{(s_2)}(R)_{\sigma_2'\sigma_2} \frac{1}{\sqrt{4\pi}}\int d\up'\,Y_\ell^m(R^{-1}\hat{\vp}')|\vp',\sigma_1'\sigma_2'\rangle \nn\\
=&\sum_{\sigma_1',\sigma_2',m'}  D^{(\ell)}(R)_{m'm} D^{(s_1)}(R)_{\sigma_1'\sigma_1} 
D^{(s_2)}(R)_{\sigma_2'\sigma_2} |\ell m',\sigma_1'\sigma_2'\rangle~.
\end{align}
In this equation we have made use of the property of the spherical harmonics 
\begin{align}
\label{280916.15}
Y_\ell^m(R^{-1}\up')=&\sum_{m'} D^{(\ell)}(R)_{m'm} Y_\ell^{m'}(\up')~.
\end{align}
Equation \eqref{280916.14}  shows that under rotation the states defined 
in Eq.~\eqref{280916.6} has the right transformation under the action of a rotation $R$, and 
 our proposition above is shown to hold.

Now, because of the transformation in Eq.~\eqref{280916.14}, corresponding to the direct product of spins 
$s_1$, $s_2$ and $\ell$, we can combine these angular momentum indices and end with the $LSJ$ basis. 
 In the latter every state is labelled by the total angular momentum $J$, 
the third component of the total angular momentum $\mu$, 
 orbital angular momentum $\ell$ and total spin $S$ (resulting from the composition of spins $s_1$ and $s_2$).
 Namely, we use the notation  $|J\mu,\ell S\rangle$ for these states which are then 
given by
\begin{align}
\label{290916.1}
|J\mu,\ell S \rangle=&\sum_{\sigma_1,\sigma_2,m,M} (\sigma_1\sigma_2 M|s_1s_2S)(m M \mu|\ell S J) |\ell m,\sigma_1\sigma_2\rangle~,
\end{align}
where we have introduced the standard Clebsch-Gordan coefficients for the composition of two angular momenta.\footnote{The Clebsch-Gordan coefficient 
$(m_1 m_2 m_3|j_1j_2j_3)$ is the  composition for $\mathbf{j}_1+\mathbf{j}_2=\mathbf{j}_3$, with $m_i$ referring to the third components of the spins.} 
Next we introduce the isospin indices $\alpha_1$ and $\alpha_2$ corresponding to the third components of the isospins 
$\tau_1$ and $\tau_2$. This does not modify any of our previous considerations 
since isospin does not transform under the action of  spatial rotations.  Within the isospin formalism the $\rho\rho$ states 
 obey Bose-Einstein statistics and these symmetric states are defined by 
\begin{align}
\label{290916.2}
|\vp,\sigma_1\sigma_2, \alpha_1\alpha_2\rangle_S=&\frac{1}{\sqrt{2}}\left(
|\vp,\sigma_1\sigma_2,\alpha_1\alpha_2\rangle + |-\vp,\sigma_2\sigma_1,\alpha_2\alpha_1\rangle
\right)~,
\end{align}
with the subscript $S$ indicating the symmetrized nature of the state under the exchange of the two particles. 
One can invert Eq.~\eqref{280916.6} and give the momentum-defined states in terms of those with well-defined orbital
 angular momentum, 
\begin{align}
\label{290916.3}
|\vp,\sigma_1\sigma_2,\alpha_1\alpha_2\rangle=&\sqrt{4\pi}
\sum_{\ell,m} Y_\ell^m(\hat{\vp})^* |\ell m,\sigma_1\sigma_2,\alpha_1\alpha_2\rangle\nn\\
=&\sqrt{4\pi}
\sum_{\scriptsize{
\begin{array}{c}
J,\mu,\ell,m\\S,M,I,t_3
\end{array}}} 
Y_\ell^m(\hat{\vp})^* (\sigma_1\sigma_2 M|s_1s_2S) (m M \mu|\ell S J) (\alpha_1 \alpha_2 t_3|\tau_1\tau_2 I)
|J\mu,\ell S,I t_3\rangle~,
\end{align}
with $I$ the total isospin of the particle pair and $t_3$ is the third component. 
Taking into account this result we can write the symmetrized states as
\begin{align}
\label{290916.4}
|\vp,\sigma_1\sigma_2,\alpha_1\alpha_2\rangle_S=&\sqrt{4\pi}
\sum_{\scriptsize{ 
\begin{array}{c}
J,\mu,\ell,m\\S,M,I,t_3
\end{array}}}
\frac{1+(-1)^{\ell+S+I}}{\sqrt{2}}(\sigma_1\sigma_2 M|s_1 s_2 S) (m M \mu|\ell S J)
 (\alpha_1 \alpha_2 t_3|\tau_1\tau_2 I) Y_\ell^m(\hat{\vp})^* |J\mu, \ell S, I t_3 \rangle~.
\end{align}
In deducing this expression we have taken into account the following symmetric properties of the Clebsch-Gordan 
coefficients
\begin{align}
\label{290916.5}
(\sigma_2\sigma_1 M|s_2 s_1 S)=& (-1)^{S-s_1-s_2}(\sigma_1\sigma_2 M|s_1 s_2 S)~,\nn\\
(\alpha_2\alpha_1 t_3|t_2 t_1 I)=&(-1)^{I-t_1-t_2}(\alpha_1\alpha_2 t_3|\tau_1\tau_2 I)~,\nn\\
Y_\ell^m(-\hat{\vp})=&(-1)^\ell Y_\ell^m(\hat{\vp})~.
\end{align}
Of course, due to the fact that we are dealing with indistinguishable bosons
 within the isospin formalism it follows that $s_1=s_2$, $\tau_1=\tau_2$ as well as that 
 $2\tau_1$ and $2s_1$ are even numbers. The combination $(1+(-1)^{\ell+S+I})/\sqrt{2}$ 
 in Eq.~\eqref{290916.4} is denoted in the following 
as $\chi(\ell S T)$ and takes into account the Bose-Einstein symmetric character of the two-particles, so that 
only states with even $\ell+S+I$ are allowed. 

The inversion of Eq.~\eqref{290916.4} gives (we assume in the following that $\ell+S+I=$even,
 so that $\chi(\ell S T)=\sqrt{2}$)
\begin{align}
\label{011016.1}
|J\mu,\ell S, I t_3\rangle = &
\frac{1}{\sqrt{8\pi}}
\sum_{\scriptsize{
\begin{array}{c}
\sigma_1,\sigma_2\\
 M,m\\
\alpha_1,\alpha_2 
\end{array}}  }
\int d\hat{\vp} Y_\ell^m(\hat{\vp}) (\sigma_1\sigma_2 M|s_1s_2S)
(m M \mu|\ell S J)
(\alpha_1 \alpha_2 t_3|\tau_1\tau_2 I)
|\vp,\sigma_1\sigma_2,\alpha_1\alpha_2\rangle_S~.
\end{align}

We can also express the state $|J\mu,\ell S,I t_3\rangle$ in terms of the states $|\vp,\sigma_1\sigma_2,\alpha_1\alpha_2\rangle$ 
without symmetrization by inverting Eq.~\eqref{290916.3}. We would obtain the same expression as Eq.~\eqref{011016.1} but 
with a factor $1/\sqrt{4\pi}$ instead of $1/\sqrt{8\pi}$, namely, 
\begin{align}
\label{061016.1}
|J\mu,\ell S, I t_3\rangle = &
\frac{1}{\sqrt{4\pi}}
\sum_{\scriptsize{
\begin{array}{c}
\sigma_1,\sigma_2\\
 M,m\\
\alpha_1,\alpha_2 
\end{array}}  }
\int d\hat{\vp} Y_\ell^m(\hat{\vp}) (\sigma_1\sigma_2 M|s_1s_2S)
(m M \mu|\ell S J)
(\alpha_1 \alpha_2 t_3|\tau_1\tau_2 I)
|\vp,\sigma_1\sigma_2,\alpha_1\alpha_2\rangle~.
\end{align}
 The extra factor of $1/\sqrt{2}$ in Eq.~\eqref{011016.1}   is a 
symmetrization factor because of the Bose-Einstein symmetry properties of the two-particle state in the symmetrized states 
$|\vp,\sigma_1\sigma_2,\alpha_1\alpha_2\rangle_S$, which disappears when employing the nonsymmetrized states.
 In order to obtain the normalization of the states $|J\mu,\ell S,I t_3 \rangle$ it is indeed simpler to use Eq.~\eqref{061016.1} though, 
of course, the same result is obtained if starting from Eq.~\eqref{011016.1}.
 The two-body particle states with definite three-momentum satisfy the normalization
\begin{align}
\label{061016.2}
\langle \vp',\sigma'_1\sigma'_2,\alpha'_1\alpha'_2|\vp,\sigma_1\sigma_2,\alpha_1\alpha_2\rangle
=&\frac{16 \pi^2 \sqrt{s}}{|\vp|} \delta(\up'-\up)~,
\end{align}
The total energy conservation guarantees that the modulus of the final and initial three-momentum in 
Eq.~\eqref{061016.3} is the same, that we denote by $|\vp|$. 
In terms of this result and Eq.~\eqref{061016.1} it follows straightforwardly by taking into account the orthogonal properties of Clebsch-Gordan 
coefficients and spherical harmonics that 
\begin{align}
\label{061016.3}
\langle  J' \mu' , \ell' S' , I' t_3' |  J  \mu , \ell S , I t_3 \rangle = &\frac{ 4 \pi \sqrt{s} }{|\vp|}
\delta_{J'J}\delta_{\mu'\mu}\delta_{\ell'\ell}\delta_{S'S}\delta_{I'I}\delta_{t'_3t_3} 
\end{align}

We are interested in the partial-wave amplitude corresponding to the transition 
between states with quantum numbers $J\bar{\ell} \bar{S} I $ to states $J \ell S I$, that 
 corresponds to the matrix element 
\begin{align}
\label{011016.2}
T^{(JI)}_{\ell S;\bar{\ell}\bar{S}}=&
\langle J\mu,\ell S,I t_3|\hat{T}|J\mu,\bar{\ell} \bar{S}, I t_3\rangle~,
\end{align}
with $\hat{T}$ the $T$-matrix scattering operator. Here we take the convention that the 
quantum numbers referring to the initial state are barred. 
Of course, the matrix element in Eq.~\eqref{011016.2} is independent of $\mu$ and $t_3$ because of 
invariance under rotations in ordinary and isospin spaces, respectively. We can calculate 
this scattering matrix element in terms of those in the basis with definite three-momentum 
by  replacing in Eq.~\eqref{011016.2} the states in the $J\ell S$ basis as given in 
Eq.~\eqref{011016.1}. We then obtain in a first step
\begin{align}
\label{021016.1}
T^{(JI)}_{\ell S;\bar{\ell}\bar{S}}=&
\frac{1}{8\pi}\sum\int d\hat{\vp}\int d\hat{\vp}' 
\,Y_\ell^m(\hat{\vp}')^* Y_{\bar\ell}^{\bar{m}}(\hat{\vp})
(\sigma_1\sigma_2 M|s_1s_2 S)(m M \mu|\ell S J)(\alpha_1\alpha_2 t_3|\tau_1\tau_2I)\nn\\
\times& (\bar{\sigma}_1\bar{\sigma}_2 \bar{M}|\bar{s}_1\bar{s}_2 S)(\bar{m} \bar{M} \mu|\bar{\ell} \bar{S} J)
(\bar{\alpha}_1\bar{\alpha}_2 t_3|\bar{\tau}_1\bar{\tau}_2I)
_S\langle \vp',\sigma_1\sigma_2,\alpha_1\alpha_2|\hat{T}|\vp,\bar{\sigma}_1\bar{\sigma}_2,\bar{\alpha}_1\bar{\alpha}_2\rangle_S~,
\end{align}
Here we have not shown the explicit indices over which the sum is done in order not to overload the notation.\footnote{They 
correspond to those indicated under the summation symbol in Eq.~\eqref{011016.1} both for the initial and final states.}
We use next the rotation invariance of the $T$-matrix operator $\hat{T}$ to simplify the previous integral so that, at the 
end, we  have just the integration over the final three-momentum angular solid. There are several steps involved 
that we  give in quite detail. The referred rotational invariance of $\hat{T}$ implies that it remains invariant 
under the transformation $\hat{T}\to R(\hat{\vp})\hat{T}R(\hat{\vp})^\dagger$, which implies at the level of the matrix elements that
\begin{align}
\label{021016.2}
_S\langle \vp',\sigma_1\sigma_2,\alpha_1\alpha_2|\hat{T}| \vp,\bar{\sigma}_1\bar{\sigma}_2,\bar{\alpha}_1\bar{\alpha}_2\rangle_S=&
_S\langle \vp',\sigma_1\sigma_2,\alpha_1\alpha_2|R(\up)\hat{T}R(\up)^\dagger| \vp,\bar{\sigma}_1\bar{\sigma}_2,\bar{\alpha}_1\bar{\alpha}_2\rangle_S~.
\end{align}
Under the action of the rotation $R(\hat{\vp})^\dagger$ ($R(\hat{\vp})^\dagger \hat{\vp}=\hat{\mathbf{z}}$ and 
$R(\hat{\vp})^\dagger \hat{\vp}'=\hat{\vp}''$) the final and initial states transform as, cf.~Eq.~\eqref{280916.13},
\begin{align}
\label{021016.3}
R(\hat{\vp})^\dagger |\vp,\bar{\sigma}_1\bar{\sigma}_2,\bar{\alpha}_1\bar{\alpha}_2\rangle_S=&\sum_{\bar{\sigma}'_1,\bar{\sigma}'_2}
D^{(\bar{s}_1)}_{\bar{\sigma}_1'\bar{\sigma}_1}(R^\dagger) D^{(\bar{s}_2)}_{\bar{\sigma}_2'\bar{\sigma}_2}(R^\dagger)
|\hat{\mathbf{z}},\bar{\sigma}'_1\bar{\sigma}'_2,\bar{\alpha}_1\bar{\alpha}_2\rangle_S~,\nn\\
R(\hat{\vp})^\dagger |\vp',\sigma_1\sigma_2,\alpha_1\alpha_2\rangle_S=&\sum_{\sigma'_1,\sigma'_2}
D^{(s_1)}_{\sigma_1'\sigma_1}(R^\dagger) D^{(s_2)}_{\sigma_2'\sigma_2}(R^\dagger)|\hat{\vp}'',\sigma'_1\sigma'_2,\alpha_1\alpha_2\rangle_S~,
\end{align}
with the convention that  $R$ inside the argument of the rotation matrices refers to $R(\hat{\vp})$.
We insert Eqs.~\eqref{021016.2} and \eqref{021016.3} into Eq.~\eqref{021016.1}, and next transform $\hat{\vp}'\to \hat{\vp}''$ as integrations variables,
 take into account the invariance of the solid angle measure under such rotation and use Eq.~\eqref{280916.15} for 
\begin{align}
\label{021016.4}
Y_{\bar{\ell}}^{\bar{m}}(\up)=&Y_{\bar{\ell}}^{\bar{m}}(R(\hat{\vp})\hat{\mathbf{z}})=
\sum_{\bar{m}'} D^{(\bar{\ell})}_{\bar{m}'\bar{m}}(R^\dagger)Y_{\bar{\ell}}^{\bar{m}'}(\hat{\mathbf{z}})~,\nn\\
Y_{\ell}^m(\up')=&Y_{\ell}^m(R(\hat{\vp})\hat{\vp}'')=\sum_{m'}D^{(\ell)}_{m'm}(R^\dagger)Y_\ell^{m'}(\hat{\vp}'')~.
\end{align}
Then, Eq.~\eqref{021016.1} for $T^{(JI)}_{\ell S;\bar{\ell}\bar{S}}$ can be rewritten as 
\begin{align}
\label{021016.5}
T^{(JI)}_{\ell S;\bar{\ell}\bar{S}}=&\frac{1}{8\pi}\sum 
 \int d\hat{\vp} \int d\hat{\vp}'' 
(\sigma_1\sigma_2 M|s_1s_2 S)(m M \mu|\ell S J) (\alpha_1\alpha_2 t_3|\tau_1\tau_2 I) 
D_{\sigma'_1\sigma_1}^{(s_1)}(R^\dagger)^* D_{\sigma'_2\sigma_2}^{(s_2)}(R^\dagger)^* \nn\\
\times& D_{m'm}^{(\ell)}(R^\dagger)^* 
Y_\ell^{m'}(\hat{\vp}'')^* 
(\bar{\sigma}_1\bar{\sigma}_2\bar{M}|\bar{s}_1\bar{s}_2 \bar{S})
(\bar{m}\bar{M}\mu|\bar{\ell}\bar{S}J)
(\bar{\alpha}_1\bar{\alpha}_2t_3|\bar{\tau}_1\bar{\tau}_2I)\nn\\
\times &
D^{(\bar{s}_1)}_{\bar{\sigma}_1'\bar{\sigma}_1}(R^\dagger)
D^{(\bar{s}_2)}_{\bar{\sigma}_2'\bar{\sigma}_2}(R^\dagger)
D^{(\bar{\ell})}_{\bar{m}'\bar{m}}(R^\dagger) 
  Y_{\bar{\ell}}^{\bar{m}'}(\hat{\mathbf{z}}) 
_S\langle \vp'',\sigma_1'\sigma_2',\alpha_1\alpha_2|\hat{T}|\,
|\vp|\hat{\vz},\bar{\sigma}_1'\bar{\sigma}_2',\bar{\alpha}_1\bar{\alpha}_2\rangle_S~.
\end{align}
Let us recall that from the composition of two rotation matrices one has that \cite{rose.021016.2,martin.290916.1}
\begin{align}
\label{021016.6}
\sum_{m_1,m_2}(m_1m_2M|\ell_1\ell_2L)D^{(\ell_1)}_{m'_1m_1}(R)
D^{(\ell_2)}_{m'_2m_2}(R)=&\sum_{M'}(m'_1m'_2M'|\ell_1\ell_2L)D_{M'M}^{(L)}(R)~.
\end{align}
We apply this result first to two combinations in Eq.~\eqref{021016.5}:
\begin{align}
\label{021016.7}
\sum_{\sigma_1,\sigma_2}(\sigma_1\sigma_2M|s_1s_2S)D_{\sigma'_1\sigma_1}^{(s_1)}(R^\dagger) 
D_{\sigma'_2\sigma_2}^{(s_2)}(R^\dagger)=&\sum_{M'}(\sigma'_1\sigma'_2 M'|s_1s_2S)D_{M'M}^{(S)}(R^\dagger)\nn\\
\sum_{\bar{\sigma}_1,\bar{\sigma}_2}(\bar{\sigma}_1\bar{\sigma}_2\bar{M}|\bar{s}_1\bar{s}_2\bar{S}) 
D^{(\bar{s}_1)}_{\bar{\sigma}'_1\bar{\sigma}_1}(R^\dagger)D^{(\bar{s}_2)}_{\bar{\sigma}'_2\bar{\sigma}_2}(R^\dagger)
=& \sum_{\bar{M}'}(\bar{\sigma}'_1\bar{\sigma}'_2\bar{M}'|\bar{s}_1\bar{s}_2\bar{S})
 D_{\bar{M}'\bar{M}}^{(\bar{S})}(R^\dagger)~,
\end{align}
so that Eq.~\eqref{021016.5} becomes
\begin{align}
\label{051016.1}
T^{(JI)}_{\ell S;\bar{\ell}\bar{S}}=&\frac{1}{8\pi}\sum\int d\hat{\vp}\int d\hat{\vp}'' 
(\sigma_1' \sigma_2' M' |s_1 s_2 S )
D_{M'M}^{(S)}(R^\dagger)^* D_{m'm}^{(\ell)}(R^\dagger)^*
(m M \mu|\ell S J) (\alpha_1\alpha_2 t_3|\tau_1\tau_2 I) Y_\ell^{m'}(\hat{\vp}'')^*\nn\\
\times & (\bar{\sigma}_1'\bar{\sigma}_2' \bar{M}'|\bar{s}_1\bar{s}_2\bar{S})
D_{\bar{M}'\bar{M}}^{(\bar{S})}(R^\dagger) D_{\bar{m}'\bar{m}}^{(\bar{\ell})}(R^\dagger)
(\bar{m}\bar{M}\mu|\bar{\ell}\bar{S}J) (\bar{\alpha}_1\bar{\alpha}_2t_3|\bar{\tau}_1\bar{\tau}_2I) 
Y_{\bar{\ell}}^{\bar{m}'}(\hat{\mathbf{z}})\nn\\
\times &
_S\langle \vp'',\sigma_1'\sigma_2',\alpha_1\alpha_2|\hat{T}|\,
|\vp|\hat{\vz},\bar{\sigma}_1'\bar{\sigma}_2',\bar{\alpha}_1\bar{\alpha}_2\rangle_S~.
\end{align}
The same relation in Eq.~\eqref{021016.6} is applied once more to the following combinations in Eq.~\eqref{051016.1}:
\begin{align}
\label{051016.2}
\sum_{m,M} (m M \mu|\ell S J) D_{M'M}^{(S)}(R^\dagger) D_{m'm}^{(\ell)}(R^\dagger)=& \sum_{\mu'}(m' M' \mu'|\ell S J) D_{\mu'\mu}^{(J)}(R^\dagger)~,\nn\\
\sum_{\bar{m},\bar{M}} (\bar{m}\bar{M}\mu|\bar{\ell} \bar{S} J) D_{\bar{M}'\bar{M}}^{(\bar{S})}(R^\dagger) D_{\bar{m}'\bar{m}}^{(\bar{\ell})}(R^\dagger)=&
\sum_{\bar{\mu}'} (\bar{m}'\bar{M}'\bar{\mu}'|\bar{\ell}\bar{S}J)D_{\bar{\mu}'\mu}^{(J)}(R^\dagger)~.
\end{align}
We take Eq.~\eqref{051016.2} into Eq.~\eqref{051016.1} which now reads
\begin{align}
\label{051016.3}
T^{(JI)}_{\ell S;\bar{\ell}\bar{S}}=&\frac{1}{8\pi}\sum  \int d\hat{\vp}\int d\hat{\vp}'' 
(\sigma_1' \sigma_2' M' |s_1 s_2 S )
(m'M'\mu'|\ell S J)
 (\alpha_1\alpha_2 t_3|\tau_1\tau_2 I) Y_\ell^{m'}(\hat{\vp}'')^*\nn\\
\times & 
(\bar{\sigma}_1'\bar{\sigma}_2' \bar{M}'|\bar{s}_1\bar{s}_2\bar{S})
(\bar{m}'\bar{M}'\bar{\mu}'|\bar{\ell}\bar{S}J)
(\bar{\alpha}_1\bar{\alpha}_2t_3|\bar{\tau}_1\bar{\tau}_2I) 
Y_{\bar{\ell}}^{\bar{m}'}(\hat{\mathbf{z}}) 
 D_{\mu'\mu}^{(J)}(R^\dagger)^*D_{\bar{\mu}'\mu}^{(J)}(R^\dagger) \nn \\
\times & 
_S\langle \vp'',\sigma_1'\sigma_2',\alpha_1\alpha_2|\hat{T}|\,
|\vp|\hat{\vz},\bar{\sigma}_1'\bar{\sigma}_2',\bar{\alpha}_1\bar{\alpha}_2\rangle_S
\end{align}
Now, the partial wave amplitude $T^{(JI)}_{\ell S;\bar{\ell}\bar{S}}$ is independent of $\mu$ so that we have that
\begin{align}
\label{051016.4}
T^{(JI)}_{\ell S;\bar{\ell}\bar{S}}=&\frac{1}{2J+1}\sum_{\mu=-J}^J T^{(JI)}_{\ell S;\bar{\ell}\bar{S}}~.
\end{align}
 The same result  in Eq.~\eqref{051016.3} is obtained with the product 
$D_{\mu'\mu}^{(J)}(R^\dagger)^*D_{\bar{\mu}'\mu}^{(J)}(R^\dagger)$ replaced by 
\begin{align}
\label{051016.5}
\frac{1}{2J+1}\sum_{\mu=-J}^J D_{\mu'\mu}^{(J)}(R^\dagger)^*D_{\bar{\mu}'\mu}^{(J)}(R^\dagger)=\frac{\delta_{\bar{\mu}'\mu'}}{2J+1}~,
\end{align}
as follows from the unitarity character of the rotation matrices.  
As a consequence any dependence in $\hat{\vp}$ present in the integrand of  Eq.~\eqref{051016.3} disappears in the 
 average of Eq.~\eqref{051016.4}, the integration in the solid angle $\hat{\vp}$ is trivial and it gives a factor $4\pi$. 
Taking into account the Kronecker delta from Eq.~\eqref{051016.5} in the third component of the total angular momentum 
and a new one that arises because $Y_{\bar{\ell}}^{\bar{m}'}(\hat{\mathbf{z}})$ is not zero only for $\bar{m}'=0$, 
we then end with the following expression for $T^{(JI)}_{\ell S;\bar{\ell}\bar{S}}$:
\begin{align}
\label{051016.6}
T^{(JI)}_{\ell S;\bar{\ell}\bar{S}}=\frac{Y_{\bar{\ell}}^{0}(\hat{\mathbf{z}})}{2(2J+1)}
\sum_{\scriptsize{\begin{array}{l} 
\sigma_1,\sigma_2,\bar{\sigma}_1\\
\bar{\sigma}_2,\alpha_1,\alpha_2\\
\bar{\alpha}_1,\bar{\alpha}_2,m
\end{array}}} &
\int d\hat{\vp}'' \, Y_\ell^{m}(\hat{\vp}'')^* (\sigma_1\sigma_2M|s_1s_2S)(m M \bar{M}|\ell S J)(\bar{\sigma}_1\bar{\sigma}_2\bar{M}| \bar{s}_1\bar{s}_2\bar{S})
(0\bar{M}\bar{M}|\bar{\ell}\bar{S}J) \nn\\
\times & (\alpha_1\alpha_2t_3|\tau_1\tau_2I)(\bar{\alpha}_1\bar{\alpha}_2t_3|\bar{\tau}_1\bar{\tau}_2I)
_S\langle \vp'',\sigma_1\sigma_2,\alpha_1\alpha_2|\hat{T}|\,
|\vp|\hat{\vz},\bar{\sigma}_1\bar{\sigma}_2,\bar{\alpha}_1\bar{\alpha}_2\rangle_S~,
\end{align}
where we have removed the primes on top of the spin and orbital angular momentum third-component symbols and 
in the previous sum $M=\sigma_1+ \sigma_2$ and $\bar{M}=\bar{\sigma}_1+\bar{\sigma}_2$.

Next, we derive the unitarity relation corresponding to our normalization for the partial-wave projected amplitudes
 $T^{(JI)}_{\ell S;\bar{\ell}\bar{S}}$. We write the $\hat{S}$ matrix as 
\begin{align}
\label{051016.7}
\hat{S}=&I-i \hat{T}
\end{align}
which satisfies the standard unitarity relation 
\begin{align}
\label{051016.8}
\hat{S} \cdot \hat{S}^\dagger=&\mathbb{I}~,
\end{align}
with $\mathbb{I}$ the identity matrix. In terms of the $T$-matrix, cf.~\eqref{051016.7}, this implies that
\begin{align}
\label{051016.9}
\hat{T}-\hat{T}^\dagger=&-i \hat{T}\hat{T}^\dagger~.
\end{align}
 Expressed with the matrix elements in the basis $\ell S J$ this relation becomes
\begin{align}
\label{051016.10}
2 {\rm Im}\hat{T}^{(JI)}_{\ell S;\bar{\ell}\bar{S}}=&- \langle J\mu,\ell S,T t_3|\hat{T}\hat{T}^\dagger |J\mu,\bar{\ell}\bar{S},T t_3\rangle~.
\end{align}
In deriving the left-hand side of this equation we have taken into account that because of time-reversal symmetry 
$T^{(JI)}_{\ell S;\bar{\ell}\bar{S}}=T^{(JI)}_{\bar{\ell}\bar{S};\ell S}$. On the right-hand side we introduce now a two-body resolution of the 
identity of states $|J\mu,\ell S,I t_3\rangle$ (we have restricted our vector space to the one generated by these states) such that, taking into account 
their normalization in Eq.~\eqref{061016.3}, one ends with
\begin{align}
\label{051016.11}
{\rm Im}T^{(JI)}_{\ell S;\bar{\ell}\bar{S}}=&-\sum_{\ell'',S''} \frac{|\vp''|}{8\pi\sqrt{s}}T^{(JI)}_{\ell,S;\ell'',S''}{T^{(JI)}}^*_{\ell'',S'';\bar{\ell}\bar{S}}~.
\end{align}
The phase space factor is included in the diagonal matrix 
\begin{align}
\label{051016.12}
\rho_{ij}=&\frac{|\vp|_i}{8\pi\sqrt{s}}\delta_{ij}~.
\end{align}
A more standard definition of the $S$-matrix implies to redefine it as
\begin{align}
\label{051016.13}
\hat{S}\to S=2\rho^{\frac{1}{2}} \hat{S} \rho ^{\frac{1}{2}}=&I-2i\rho^{\frac{1}{2}}\hat{T}\rho^{\frac{1}{2}}~,
\end{align}
such that now the diagonal matrix elements of the identity operator $I$ and $S$ are just 1 and $\eta_ie^{2i\delta_i}$, in order, 
where $\eta_i$ is the inelasticity for channel $i$  and $\delta_i$ its phase shift.


\end{document}